\documentclass[sigconf]{acmart}
\providecommand{\tagpdfsetup}[1]{}
\usepackage{multirow}
\usepackage{subcaption}
\usepackage[table]{xcolor}

\AtBeginDocument{%
  }

\definecolor{linkblue}{RGB}{0,102,204}
\setcopyright{none}
\renewcommand\footnotetextcopyrightpermission[1]{}
\makeatletter
\patchcmd{\maketitle}{\@mkabstract}
  {\noindent\begin{minipage}[t][\dimexpr\@colht-\topskip\relax][t]{\columnwidth}%
   \vspace{0pt}\@mkabstract}
  {}{\PackageError{midas-layout}{Cannot begin abstract column}{Check the acmart version.}}
\patchcmd{\maketitle}{\let\metadata@authors=\authors}
  {\end{minipage}\par\let\metadata@authors=\authors}
  {}{\PackageError{midas-layout}{Cannot end abstract column}{Check the acmart version.}}
\patchcmd{\maketitle}{\@specialsection{\keywordsname}}
  {\par\nobreak\smallskip\noindent
   {\bfseries\keywordsname}\par\nobreak}
  {}{\PackageError{midas-layout}{Cannot format keywords}{Check the acmart version.}}
\makeatother
\copyrightyear{2027}
\acmYear{2027}
\acmDOI{XXXXXXX.XXXXXXX}
\acmConference[CHI '27]{Proceedings of the 2027 CHI Conference on Human Factors in Computing Systems}{May 10--14, 2027}{Pittsburgh, USA}
\acmBooktitle{Proceedings of the 2027 CHI Conference on Human Factors in Computing Systems (CHI '27), May 10--14, 2027, Pittsburgh, USA}
\acmISBN{978-1-4503-XXXX-X/26/10}
\acmSubmissionID{9558}

\usepackage{tikz}
\usetikzlibrary{arrows.meta, positioning}

\begin{document}

\title{MidasTUI: Instant Tangible User Interfaces for Wearable Displays using Everyday Objects}

\author{Wen Ying}
\orcid{0000-0001-8230-1920}
\affiliation{
    \department{Department of Computer Science}
    \institution{University of Virginia}
    \city{Charlottesville}
    \state{Virginia}
    \country{USA}
}
\email{wenying@virginia.edu}

\author{Xuweiyi Chen}
\orcid{0009-0009-4426-5637}
\affiliation{
    \department{Department of Computer Science}
    \institution{University of Virginia}
    \city{Charlottesville}
    \state{Virginia}
    \country{USA}
}
\email{xuweic@virginia.edu}

\author{Zezhou Cheng}
\orcid{0000-0001-7754-0871}
\affiliation{
    \department{Department of Computer Science}
    \institution{University of Virginia}
    \city{Charlottesville}
    \state{Virginia}
    \country{USA}
}
\email{zc3bp@virginia.edu}

\author{Seongkook Heo}
\orcid{0000-0003-2004-4812}
\affiliation{
  \department{Department of Computer Science and Engineering}
  \institution{Ulsan National Institute of Science and Technology}
  \city{Ulsan}
  \country{South Korea}}
\email{seongkook@unist.ac.kr}

\renewcommand{\shortauthors}{Ying et al.}

\begin{abstract}
While wearable displays can bring digital workspaces anywhere, interaction remains limited by mid-air gestures that lack tactile feedback and induce fatigue, or controllers that are cumbersome to carry.
We present MidasTUI, a system that instantly transforms arbitrary everyday objects into opportunistic tangible user interfaces through brief interaction demonstrations. For example, a user can slide an earbud case across a table as a mouse, enabling opportunistic interaction without carrying a dedicated input device.
MidasTUI detects when a user is holding and moving an object using the wearable display camera and a Transformer-based temporal classifier. It then infers the intended input modality (e.g., slider, mouse, keypad) from hand movement and maps the tracked hand-object interaction to an application function.
Our evaluations demonstrate reliable cross-object and cross-user generalization. A comparative user study further shows that grounding interaction in everyday objects reduces physical burden and improves perceived usability and control relative to mid-air interaction.

\end{abstract}



\begin{teaserfigure}
  \centering
  {\fontsize{10}{12}\selectfont
    Project page: \href{https://midastui.github.io/}{\textcolor[HTML]{455fd4}{https://midastui.github.io/}}}\\[6pt]
  \Description[Four-panel overview showing how everyday objects are demonstrated, assigned input behaviors, and mapped to application functions.]
  {The left panel shows a seated headset user interacting with everyday objects on a table: sliding an earbud case, rotating a candy box, and tapping a book. Colored callouts label these actions as mouse, knob, and keypad interactions. The second panel illustrates the three demonstrations separately, showing sliding, rotating, and tapping motions. The third panel shows behavior-based input inference, where the system determines an input type from the demonstrated hand-object motion and assigns the corresponding control behavior to each object: the earbud case becomes a mouse, the candy box becomes a knob, and the book becomes a keypad. The right panel shows these tangible inputs mapped to functions in a wearable photo-editing interface, including cursor control, value adjustment, and tool selection.}
    \includegraphics[width=\textwidth]{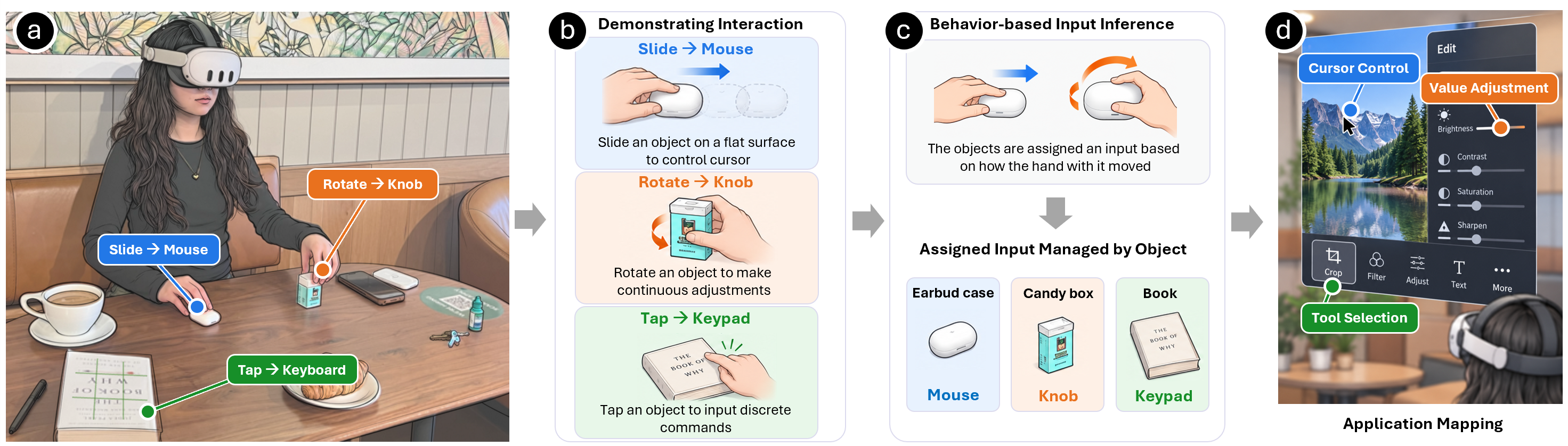}
  \caption{MidasTUI enables users to turn everyday objects into tangible controls for wearable displays by demonstrating how they want to use them. (a-b) A user slides, rotates, or taps available objects, and (c) the system infers the intended input type (e.g., mouse, knob, or keypad) from hand–object motion and assigns the corresponding control to each object. (d) These tangible inputs are then mapped to application functions, enabling users to work productively across everyday environments.}
  \label{fig:teaser}
\end{teaserfigure}

\maketitle

\fancyhead{}
\fancyhead[L]{\sffamily\footnotesize\shorttitle}
\fancyhead[R]{\sffamily\footnotesize\shortauthors}

\newpage 
\vspace*{0pt} 
\section{Introduction}
\label{sec:intro}

Wearable displays support productivity beyond the traditional desktop by bringing virtual applications into ``on the go'' everyday environments~\cite{grubert2019office, simeone2023everyday}. However, interaction with these displays using spatial interfaces still largely depends on gaze~\cite{lystbaek2022gaze, isomoto2022dwell} and mid-air gestures~\cite{Koutsabasis08112019, ying2026redirected}, which often suffer from limited precision and physical fatigue due to the lack of tangibility and haptic feedback~\cite{hinckley1994survey, hansberger2017dispelling, hirzle2020survey}. Dedicated input devices, such as controllers~\cite{souchet2023narrative, luong2023controllers} and mice~\cite{Guo2019OfficeVR}, as well as companion devices such as tablets~\cite{biener2020breaking, shi2025reachpad} can provide more accurate and comfortable control, but at the cost of portability and battery life, and they may not always be readily available.

To provide more convenient yet accurate and comfortable control, researchers explored turning everyday physical objects into tangible user interfaces (TUIs)~\cite{corsten2013instant, hettiarachchi2016annexing, fujinami2018painting, zhou2020gripmarks, drogemuller2021turning, du2022opportunistic, gong2023affordance, monteiro2023teachable, liang2025physical}, leveraging their ready availability and inherent tangibility to support effective spatial interaction~\cite{ishii1997tangible, terrenghi2007affordances}. Tangible objects also support asymmetric bimanual interaction, where each hand can be assigned to a different function~\cite{lyu2026objectures,ying2026physical}. This division of labor between the hands has been shown to improve efficiency in compound tasks~\cite{guiard1987asymmetric, two1994kabbash, Hinckley1997Bimanual, webb2016wearables} but difficult to achieve with mid-air gestures, which largely afford symmetric input. However, most of these TUI interactions are object-centric and require external sensing devices~\cite{fan2018what, yeo2018exploring, rudolph2022sensing, chen2024viobject} or extensive upfront setup~\cite{drogemuller2021turning}, in which each object's geometric~\cite{sheridan2006affordance, henderson2008opportunistic, he2024AdapTUI}, texture~\cite{fujinami2018painting}, kinetic capability~\cite{zhou2020gripmarks, gong2023affordance}, or semantic information~\cite{corsten2013instant, liang2025physical} must be pre-registered and bound to predetermined digital controls before use~\cite{drogemuller2021turning, he2023ubi, monteiro2023teachable}. This requirement makes such systems less suited to ad-hoc, everyday scenarios, since users cannot simply pick up an unfamiliar object in a new environment and begin interacting, and the bindings may transfer poorly across objects whose shapes and affordances were not anticipated at design time~\cite{gong2023affordance}. In addition, many existing TUI systems focus on design exploration~\cite{gong2023affordance, liang2025physical}, rapid prototyping~\cite{kelly2018arcadia}, or conceptual demonstrations~\cite{heun2013smarter, du2022opportunistic}, with less emphasis on real-time usable systems for sustained productivity scenarios such as work on wearable displays. 
These limitations mean that prior systems remain confined to selected object sets and pre-authored scenarios, falling short of the truly opportunistic, ``pick up anything and use it'' interaction that wearable displays demand in everyday environments.

In this work, we present MidasTUI, a novel hand-centric approach that transforms available everyday objects into opportunistic tangible user interfaces without extensive configuration. Unlike previous object-centric TUI systems that rely on specific objects or pre-authored bindings, MidasTUI adapts to whatever objects are at hand, allowing the user to simply grab an object and manipulate it as they would a familiar input device~\cite{harrison2014touchtools}.
For example, a user can pick up an earbud case and slide it across a coffee table as a mouse, enabling precise and comfortable interaction (Figure~\ref{fig:teaser}). MidasTUI also extends naturally to bimanual use, where the user can hold a different object in each hand to drive distinct functions in parallel.
At a high level, MidasTUI detects when the user is holding and moving an object using the wearable display camera, infers the intended input modality (e.g., slider, mouse, keypad) from the user's hand movement, and maps the tracked interaction to an application function.

Rather than relying on predefined mappings or explicit configuration, MidasTUI treats user demonstration as a mechanism for defining how objects function as input. This approach operates through a four-stage pipeline and was tested with 16 participants. First, \emph{Hand–Object Engagement Detection (the Move Together model)} uses a FastViT spatial backbone with a Transformer temporal head to determine whether the user's hand grabs and moves a nearby object (F1=0.92 across 20~objects, < 6 ms per inference). Second, \emph{Input Modality Inference (the Summon model)} infers the intended input modality from hand movement: a generalized model provides immediate top-3 suggestions without per-user data (80.9\% top-1 across 10~classes), and a personalized model adapts to the user's motion style, reaching 97.5\% accuracy. Third, the \emph{TUI Manager} stores and manages recognized TUIs and their interaction models. Lastly, \emph{Interaction-to-Application Function Mapping} binds each TUI interaction to an application function according to its dimensionality (1D/2D/3D), motion primitives (linear/rotary), flow (continuous, discrete), and control mapping (absolute/relative). Beyond technical feasibility, we also investigate whether interaction with everyday objects provides practical benefits over bare-hand mid-air input. We recruited 10 participants to compare MidasTUI with Mid-Air interaction in an exploratory photo-editing task involving tool switching, parameter adjustment, and object manipulation in MR. Although MidasTUI introduced additional acquisition and interaction overhead, participants reported better usability and control, and highlighted less physical fatigue, stable spatial references, and useful bimanual interaction as key advantages.

In summary, this work contributes: (1) a demonstration-driven approach that enables users to define input behavior for everyday objects through interaction, supporting opportunistic tangible interaction on wearable displays; (2) two novel models, \emph{Move Together} for hand–object engagement detection and \emph{Summon} for input modality inference, along with the corresponding datasets we collected to train and evaluate them; (3) technical evaluations showing robust performance across a diverse set of everyday objects; and (4) A user study comparing MidasTUI with mid-air interaction in a realistic photo-editing task, showing that everyday-object tangibility reduces perceived physical burden and improves usability and control.
\section{Related Work}
\label{sec:background}

\subsection{Ad-hoc Interaction for Wearable Displays}
\label{sec:bg-wearble}

With the development of head-mounted displays (AR/VR) and smart glasses, wearable displays are emerging as a promising platform for daily work and leisure tasks, offering mobile, extendable, and ``on the go'' workspaces that adapt to user needs and contexts~\cite{grubert2019office, ofek2020towards, simeone2023everyday, cai2025aiget}. However, providing input that is both readily available in everyday contexts and efficient enough for sustained use remains challenging.

Prior work has explored ad-hoc input techniques by leveraging the user's body, including gaze~\cite{lystbaek2022gaze, isomoto2022dwell, wagner2023fitts}, head movement~\cite{yan2018headgesture, le2021vxslate} and mid-air gestures~\cite{Koutsabasis08112019, tian2025summonbrush, ying2026redirected}. These techniques are always available but often suffer from fatigue, low precision, lack of haptic feedback, and limited expressiveness~\cite{hansberger2017dispelling, luong2023controllers, hirzle2020survey}. Other work has leveraged dedicated input devices such as controllers~\cite{souchet2023narrative, luong2023controllers} and mice~\cite{Guo2019OfficeVR}, companion devices such as tablets~\cite{biener2020breaking, shi2025reachpad}, and nearby grounded surfaces~\cite{xiao2018mrtouch, medeiros2023benefits, dupre2024tripad}. They can provide more accurate and comfortable control through tangible affordance and physical support~\cite{Arora2017SketchingInVR, cheng2022comfortable, medeiros2023benefits, ying2026physical}. However, dedicated devices must be charged and carried, and grounded surfaces lack portability, neither of them is always readily available in everyday environments.

Our work instead explores how everyday objects can be appropriated as input devices through user demonstration, combining the availability with the ability to flexibly define their input behavior. 

\begin{figure*}[t]
  \centering
  \Description[Four-stage pipeline diagram showing Move Together detection, Summon classification, interaction management, and application mapping.]{A flow diagram illustrating the MidasTUI pipeline. Stage one: the Move Together classifier processes passthrough camera frames to detect hand-object engagement. Stage two: the Summon phase records a user demonstration and classifies it into one of ten input modalities. Stage three: the interaction manager tracks the object state. Stage four: the inferred interaction is mapped to virtual interface controls in the active application.}
  \includegraphics[width=1\linewidth]{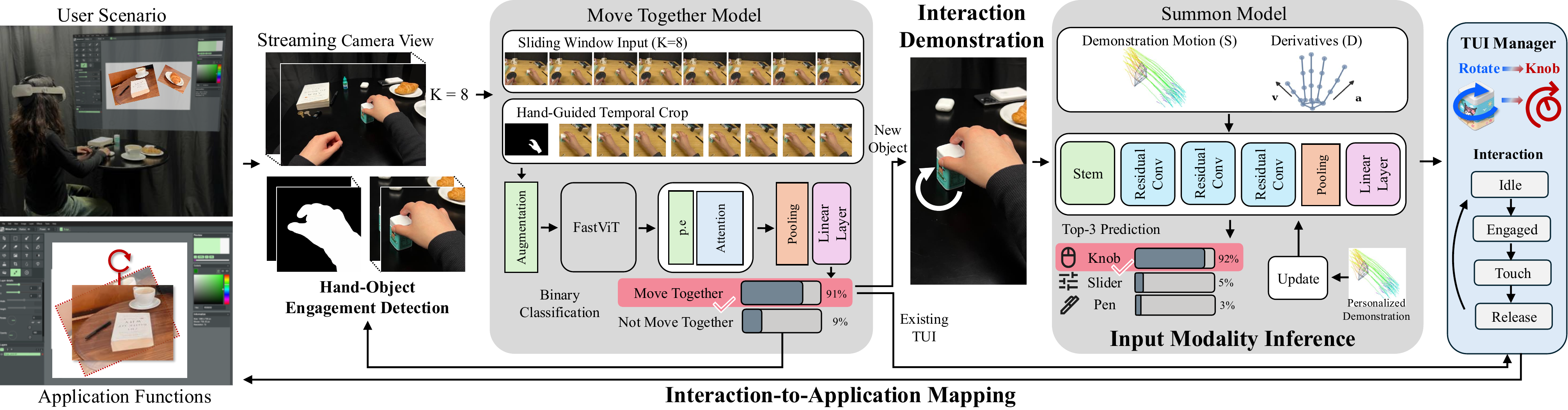}
  \vspace{-4mm}
  \caption{MidasTUI pipeline overview. The system detects and monitors hand--object engagement using the \emph{Move Together} classifier. When engagement is detected for a new object, the user enters the \emph{Summon} phase to demonstrate an interaction, which is classified into an input modality and mapped to the active application.}
  \vspace{-3mm}

  \label{fig:pipeline-overview}
\end{figure*}

\subsection{Utilizing Everyday Objects as Tangible User Interfaces}
\label{sec:bg-tui}

Tangible user interfaces (TUIs) bind physical objects to digital controls, leveraging the user's familiarity with manipulating real-world artifacts to support efficient and embodied interaction~\cite{fitzmaurice1995bricks, ishii1997tangible}. Beyond providing passive haptic feedback, tangible objects also support asymmetric bimanual interaction, where each hand can be assigned to a different function~\cite{bier1993toolglass, lyu2026objectures}. Such coordination improves efficiency and reduces completion time and cognitive effort in compound tasks~\cite{guiard1987asymmetric, two1994kabbash, Hinckley1997Bimanual, webb2016wearables}, 
but is hard to achieve with mid-air gestures, which lack the stable physical reference and passive haptic feedback that asymmetric input requires.

A large body of work has explored repurposing everyday objects as TUIs, motivated by their ubiquity, low cost, and rich physical affordances~\cite{terrenghi2007affordances}. Some research takes a \emph{user-authored} approach, letting users explicitly bind objects to digital functions through demonstration, voice commands, or interactive teaching~\cite{corsten2013instant, drogemuller2021turning, wang2021gesturar, he2023ubi, monteiro2023teachable}. While effective for customization, these systems place the burden of configuration on end-users, making them unsuitable for spontaneous, ad-hoc interaction in dynamic everyday environments.

Other systems pursue automatic object-function binding, which can be further grouped into two strategies. \emph{Object-centric} methods map specific physical properties of an object, such as geometry~\cite{henderson2008opportunistic, gong2023affordance, he2024AdapTUI}, texture~\cite{fujinami2018painting}, kinetic capability~\cite{zhou2020gripmarks, gong2023affordance}, or semantic category~\cite{corsten2013instant, liang2025physical}, to predetermined interface roles. However, they typically rely on external sensing hardware~\cite{fan2018what, yeo2018exploring, rudolph2022sensing, chen2024viobject} or extensive upfront setup~\cite{drogemuller2021turning} to register each object's properties before use. This requirement makes such systems less suited to ad-hoc, everyday scenarios, since users cannot simply pick up an unfamiliar object in a new environment and begin interacting. In addition, object-centric systems depend on the ``right'' object being available, may adapt poorly to user intent, and often transfer poorly to objects whose shapes and affordances were not anticipated at design time~\cite{gong2023affordance}.

In contrast, \emph{hand-centric} methods recognize tool-use gestures from the hand alone, generalizing across whichever object is held~\cite{harrison2014touchtools, sharma2019grasping, zhou2020gripmarks, sharma2024graspui}. However, existing hand-centric systems rely on pre-defined gesture vocabularies that may not reflect how people naturally handle objects. As a result, interaction semantics are determined by the system designer rather than emerging from the user's own manipulation behavior. 
Moreover, much of this work remains oriented toward design exploration~\cite{gong2023affordance, liang2025physical}, rapid prototyping~\cite{kelly2018arcadia}, or conceptual demonstrations~\cite{heun2013smarter, du2022opportunistic}, rather than sustained, real-time use in productivity scenarios.

Our work builds on \emph{user-authored} and \emph{hand-centric} approaches, but takes a different perspective on how interaction should be defined. Rather than treating demonstration as a configuration step or relying on predefined gesture vocabularies, we treat user demonstration as a mechanism for defining interaction semantics. In our approach, the system interprets how users manipulate objects as an expression of intended input modality, enabling real-time, object-independent interface creation without explicit configuration. This shifts the role of everyday objects from predefined interface elements to flexible interaction substrates, whose function emerges from user action rather than being pre-authored.

\subsection{Hand--Object Interaction Detection}
\label{sec:bg-hoi}

A prerequisite for enabling opportunistic tangible interaction with everyday objects is the ability to reliably detect when a user is holding and manipulating an object, and to infer the intended input from that interaction. Prior work has approached this problem with instrumented sensing, including smartwatches~\cite{leung2018twistin, chen2024viobject} and customized wearable sensors~\cite{grosse2017finding, fan2018what, yeo2018exploring, rudolph2022sensing}, which provide reliable signals at the cost of additional hardware. To support uninstrumented use, camera-based approaches detect hand--object contact~\cite{fujinami2018painting, mollyn2024egotouch} or recognize predefined manipulation gestures~\cite{sharma2019grasping, zhou2020gripmarks, sharma2024graspui}. However, robust camera-based detection in unconstrained, in-the-wild settings remains challenging due to occlusion, object diversity, and the fine-grained nature of hand--object contact.

Existing datasets of camera-based hand--object interaction~\cite{brahmbhatt2020contactpose, kwon2021h2o} primarily target functional grasp recognition, capturing how the hand contacts an object to perform a specific task such as pouring or opening. These benchmarks are not directly applicable to our setting, which goes beyond functional purpose and instead focuses on whether the hand and object are moving together. We therefore collect our own \emph{Move Together} dataset, on which we train a detector that identifies \emph{correlated motion} between the hand and a nearby object. This signal requires no object instrumentation or predefined gesture vocabulary, and generalizes across object shape, size, and material, allowing engagement to be treated as an object-independent interaction cue.

Once engagement is detected, the system must further infer \emph{what kind of input} the user is performing. Prior work shows that fine-grained hand motion features can be modeled to recover user intent for tasks such as grasp-onset prediction~\cite{valkov2025grasp}, selection-intent detection~\cite{wang2022rids}, ray-landing prediction~\cite{xu2025predicting}, and dynamic gesture recognition~\cite{lu2016dynamic, park2015virtual}, typically using temporal models over kinematic features (joint angles, velocity, acceleration). However, these methods classify within a fixed vocabulary tied to a specific task. Our \emph{Summon} model adopts similar temporal-modeling principles but instead infers the \emph{input modality} the user is demonstrating (e.g., mouse, slider, knob). This shifts input definition from a fixed, system-defined vocabulary to a user-driven process, where interaction semantics are inferred from how users demonstrate their intended behavior.

\section{MidasTUI System}
\label{sec:design}

MidasTUI enables tangible interaction on wearable displays using everyday objects, without requiring dedicated input devices. It is a camera-based, hand-centric, user-authored system organized into four stages: (1) \emph{Hand-Object Engagement Detection}, (2) \emph{Input Modality Inference}, (3) \emph{Interaction Tracking}, and (4) \emph{Interaction-to-Application Function Mapping}. We describe each stage below.

\subsection{Hand–Object Engagement Detection Model}
\label{sec:design-pick}

\subsubsection{Interaction Principle}
The first stage of the MidasTUI pipeline determines whether the user intends to use an physical object as an input modality for interacting with the wearable display. Rather than requiring an explicit selection gesture (e.g., ray-cast or gaze dwell) or a dedicated input device (e.g., mouse or touchpad), we infer engagement implicitly: when a user grasps an object and begins to move it, the correlated motion of the hand and the object signals interaction intent. We call this the hand--object engagement detection model, or \textbf{Move~Together} for short. The system continuously monitors the scene through the headset's passthrough cameras. For each frame, it identifies whether the hand and the object exhibit correlated motion through a \emph{move~together} classifier. If so, the system concludes that the user intends to engage with the object, has picked it up, and activates the MidasTUI pipeline for that object; otherwise, no TUI interaction is registered (e.g., when the hand hovers over or rests on the object without displacing it).

\subsubsection{Move Together Classifier}
We frame engagement detection as a binary temporal classification task. Let $\mathbf{X} = (I_t, \dots, I_{t+K-1})$ denote a sliding window of $K$ consecutive frames from the passthrough camera, where each frame $I_k \in \mathbb{R}^{3 \times H \times W}$ is an RGB image cropped around the hand--object contact area and resized to $H{=}W{=}224$. The classifier predicts a label $y = f(\mathbf{X}) \in \{0, 1\}$, where $y{=}1$ indicates that the hand and a nearby object are moving together (engagement), and $y{=}0$ indicates incidental or no correlated motion. We decompose $f$ into a spatial encoder $g$ and a temporal aggregation head $h$: $f(\mathbf{X}) = \sigma\!\bigl(h(g(\mathbf{X}))\bigr)$, where $\sigma$ is the sigmoid function.

\subsubsection{Model Architecture}
Each frame is independently encoded by a FastViT-T8~\cite{vasu2023fastvit} backbone $g$. Global average pooling produces a $D$-dimensional backbone feature vector ($D{=}384$), yielding a temporal feature sequence
\[
  \mathbf{Z} = (g(I_t), \dots, g(I_{t+K-1})) = (\mathbf{z}_1, \dots, \mathbf{z}_K) \in \mathbb{R}^{K \times D}.
\]

The temporal head maps $\mathbf{Z}$ to a single scalar logit. We compare four designs: Mean, TCN~\cite{bai2018empirical}, Transformer~\cite{vaswani2017attention}, and BiLSTM~\cite{graves2005framewise}. The Transformer and BiLSTM heads project each 384-dimensional frame feature to 256 dimensions before temporal modeling; the Mean and TCN heads also use 256-dimensional hidden layers. The Mean head averages frame features across time,
\[
  h_{\text{mean}}(\mathbf{Z}) = \mathrm{MLP}\!\left(\frac{1}{K}\sum_{k=1}^{K}\mathbf{z}_k\right),
\]
and serves as a position-invariant baseline. The TCN head applies temporal convolutions to capture local motion patterns. The BiLSTM head captures sequential dynamics in both forward and backward directions. The Transformer head uses positional encodings and self-attention to model longer-range temporal dependencies. All heads output a single logit, which is passed through $\sigma$ to produce the binary prediction $\hat{y}_i = \sigma\!\bigl(h(g(\mathbf{X}_i))\bigr)$.

The Mean, TCN, and Transformer heads operate on fixed-length $K$-frame windows sampled with stride~1. The BiLSTM head processes overlapping chunks of $K$ frames with stride $K/2$, producing per-frame predictions that are averaged over each chunk. We ablate the window size $K \in \{8, 16, 32\}$ in Section~\ref{sec:eval-move-together}.

\subsubsection{Training Setup}
All models are trained end-to-end by minimizing the binary cross-entropy loss over $N$ training clips:
\[
  \mathcal{L} = -\frac{1}{N}\sum_{i=1}^{N}
  \bigl[
    y_i \log \hat{y}_i + (1 - y_i) \log(1 - \hat{y}_i)
  \bigr],
\]
where $y_i \in \{0,1\}$ is the ground-truth label and $\hat{y}_i = \sigma\!\bigl(h(g(\mathbf{X}_i))\bigr)$ is the predicted probability for clip $\mathbf{X}_i$.

Each frame is cropped around the headset-tracked hand mask, padded by 30\% around the bounding box, and resized to \(224\times224\). The crop is recomputed for every frame in a chunk so the model stays focused on one hand despite camera, body, or background motion. We apply random horizontal flip (\(p=0.5\)) and bounded color jitter, with augmentation sampled once per chunk so all frames in a clip share the same transform. Although training uses only single-hand sequences, the mask-guided crop always shows one tracked hand, and random flipping removes handedness bias. At deployment, the bimanual extension (Section~\ref{sec:bimanual}) runs the same network twice in parallel, once per hand mask, and treats the outputs as independent move-together signals.

\begin{figure*}[t]
  \centering
  \Description[Ten input device classes with example objects and corresponding hand motion trajectories.]{A two-row figure showing ten input device classes used in Summon data collection. Top row: photographs of physical objects mimicking each device type (slider, knob, switch, mouse, touchpad, pen, joystick, d-pad, keypad, pointer). Bottom row: corresponding 3D hand motion trajectory visualizations showing the characteristic movement pattern for each class.}
  \includegraphics[width=1\linewidth]{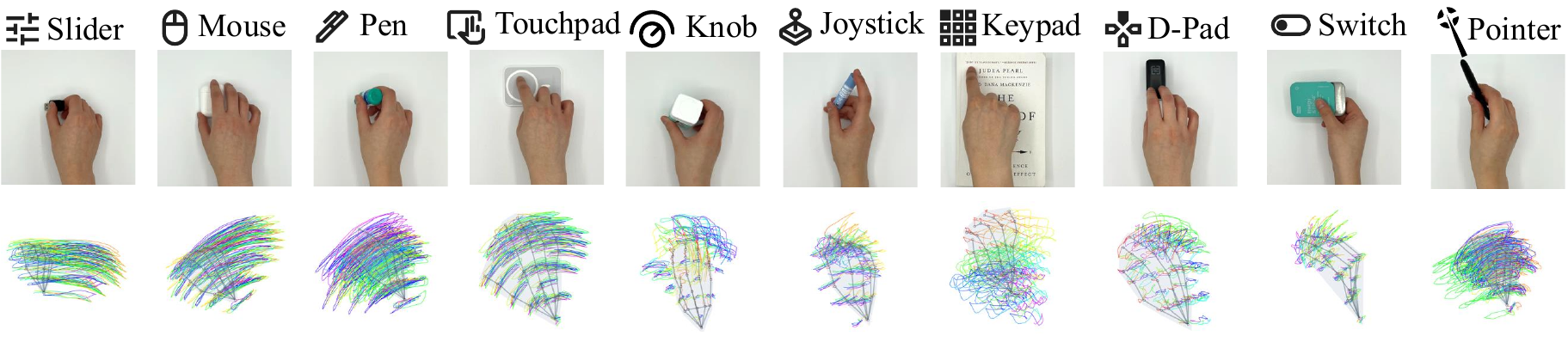}
  \vspace{-5mm}
  \caption{We collected ``summon'' data using 10 common input devices. For each device, the interaction was mimicked by touching or moving a corresponding physical object, and the bottom row shows the resulting motion trajectories.\vspace{-2mm}}

  \label{fig:summon-objects}
\end{figure*}

\subsection{Input Modality Inference Model}
\label{sec:design-summon}

\subsubsection{Interaction Principle}
Once the Move~Together detector confirms that the user has engaged with a physical object, the system enters the input modality inference stage, which we call \textbf{Summon}. In this stage, the user effectively ``summons'' an interface by demonstrating how they intend to use the object as an input device. The key idea is that everyday objects afford different interaction styles: for example, a soap bar can be slid like a mouse, while a bottle cap can be twisted like a dial. The user's natural demonstration of these movements therefore reveals the interaction modality they intend to invoke.

MidasTUI adopts a two-stage recognition strategy. A \emph{generalized model}, trained on data from other users, provides immediate predictions without requiring per-user calibration. It presents the top three most likely interaction modalities for the user to choose from. As the user makes selections over time, these choices become labeled examples for training a \emph{personalized model} adapted to that individual's motion style, progressively improving accuracy from 80.9\% to 97.5\%, which we discussed in Section~\ref{sec:eval-summon}.

\subsubsection{Interaction Demonstration}
During the Summon stage, the user holds the engaged object and performs the motion they wish to use for input. For example, a user might slide a pendrive back and forth to indicate slider control, or rotate a pill box to indicate rotary input (Figure~\ref{fig:summon-objects}). The system records the 3D hand motion throughout this demonstration using the headset's built-in hand tracking, capturing the world-space positions of 26 hand joints at each timestep.

\subsubsection{Input Modality Prediction}
\label{sec:input-prediction}

We formulate input modality inference as a multiclass temporal classification task: given a variable-length demonstration sequence, the model predicts which of $C=10$ input device classes the user is mimicking.

\paragraph{Input Representation.}
Let the demonstration sequence be
$\mathbf{S} = (\mathbf{s}_1, \mathbf{s}_2, \ldots, \mathbf{s}_T)$,
where each frame $\mathbf{s}_t \in \mathbb{R}^{78}$ contains the $(x, y, z)$ world-space coordinates of all 26 tracked hand joints. To capture motion dynamics beyond static hand pose, we augment each frame with first- and second-order temporal derivatives:
\[
  \mathbf{v}_t = \mathbf{s}_t - \mathbf{s}_{t-1}, \qquad
  \mathbf{a}_t = \mathbf{v}_t - \mathbf{v}_{t-1},
\]
where $\mathbf{v}_t$ and $\mathbf{a}_t$ denote per-frame velocity and acceleration, respectively. Concatenating these features yields an enriched representation $\mathbf{s}_t' = [\mathbf{s}_t \mathbin\Vert \mathbf{v}_t \mathbin\Vert \mathbf{a}_t] \in \mathbb{R}^{234}$ for each frame.

\paragraph{Classification Model.}
We use a 1D residual convolutional network (ResNet1D) that operates directly on the temporal feature sequence $\mathbf{S}' \in \mathbb{R}^{T \times 234}$. The architecture consists of a stem convolution with kernel size 7, followed by three downsampling residual stages that progressively increase the channel dimension while reducing temporal resolution through stride-2 convolutions. A length-aware masked global average pooling layer aggregates the variable-length sequence, and a final fully connected layer produces a probability distribution over $C=10$ input device classes (Figure~\ref{fig:summon-objects}): \emph{slider}, \emph{knob}, \emph{switch}, \emph{mouse}, \emph{touchpad}, \emph{pen}, \emph{joystick}, \emph{d-pad}, \emph{keypad}, and \emph{pointer}. Together, these classes span continuous and discrete interaction flows, linear and rotary motion primitives, one- to three-dimensional control, and both absolute and relative mappings.
Architecture details for both classifiers are provided in Appendix~\ref{appendix:arch}.

\subsubsection{Training and Adaptation}
We train a generalized model for immediate use and adapt it to individual users as labeled demonstrations accumulate.

\paragraph{Training.}
We use a 1D residual convolutional network adapted from ResNet~\cite{he2016deep} to model temporal hand-motion features.
The model is trained end-to-end using cross-entropy loss~\cite{goodfellow2016deep}. We further apply Mixup regularization~\cite{zhang2017mixup}, which constructs virtual training examples by linearly interpolating pairs of input sequences and labels.

Data augmentation includes random temporal cropping, Gaussian noise injection, scale jitter, and random time reversal.
Training configurations for both classifiers are provided in Appendix~\ref{appendix:training}.

\paragraph{Generalized Model.}
The generalized model is trained on demonstrations from all participants except the target user and evaluated using leave-one-participant-out cross-validation over 16 participants. At inference time, the model returns a ranked list of the top three most probable interaction classes, from which the user selects the intended modality.

\paragraph{Personalized Model.}
Each user selection serves as a new labeled training example. After 2--3 demonstrations per class have been collected through natural use, the system fine-tunes a personalized model on this user-specific data. By observing user's motion style during training, the personalized model can better account for cross-user variability that limits the generalized model (Section~\ref{sec:eval-summon}). Once the input modality is determined, the system transitions to interaction mode, in which the object's movements are continuously interpreted as input events for the active application.

\subsection{TUI Manager}
\label{sec:tui-manager}
Following hand-object engagement detection and input modality inference, the TUI Manager governs how an engaged object is activated, recognized across sessions, and tracked during runtime control. It is built around three coupled mechanisms: (i) an interaction area calibration that defines where TUI control is valid, (ii) a TUI registry that lets previously inferred objects be re-summoned without re-demonstration, and (iii) a state machine that converts hand-object motion into digital control events for two hands.

\subsubsection{Interaction Area Calibration}
MidasTUI restricts TUI activation to a user-calibrated \emph{interaction area} on arbitrary tabletop. At setup, the user defines the area by pointing the index fingertip at four corners in clockwise order. At runtime, the system maps a 3D hand position onto the quadrilateral interaction area via inverse bilinear interpolation, producing normalized coordinates $(u,v) \in [0,1]^2$. A control–display (CD) gain then maps relative displacements within this physical region to cursor displacements traversing the full screen.

Only objects manipulated within this area are treated as potential TUIs. When an object moves inside the region, the system checks whether it has already been registered and, if so, activates its associated interaction. Manipulations outside the area are treated as normal everyday interactions and do not trigger TUI behavior. This spatial filtering reduces unintended activation by preventing the system from interpreting every touched object as an input device, while preserving everyday interaction habits by allowing the same object to serve both ordinary physical and digital control functions depending on where it is used. For example, a cup manipulated within the interaction area may be interpreted as a knob control for parameter adjustment, while outside the area it can still be used for drinking, enabling seamless transitions between instrumental and everyday object use.

\subsubsection{TUI Registration and Recognition}
After the system detects hand-object engagement for an object manipulated within the interaction area, it first determines whether that object has already been assigned an input (Figure~\ref{fig:pipeline-overview}), as  MidasTUI maintains a list of objects registered as TUIs for instant and repeated use.  
If the object has not yet been registered, the system proceeds to the second stage, \emph{input modality inference}, to infer the intended input modality. The object is then registered as a new TUI and added to the registered TUI list. If the object has already been registered, the system skips the inference stage and directly enters the \emph{interaction tracking} stage. This allows previously inferred objects to be reused with their existing control mappings, avoiding redundant inference and reducing interaction overhead.

To recognize whether an engaged object has already been registered, the system crops the camera view around the hand after hand-object engagement is detected. The cropped hand-object view is then compared against the list of registered hand-object views. We measure object similarity using DINO~\cite{caron2021emerging}, which provides high-level visual embeddings that capture semantic visual similarity between images. High similarity scores indicate a match to a previously registered object. Based on this similarity, the system determines whether the current object corresponds to an existing TUI or should be treated as a newly encountered object. Specially, if the best match exceeds $0.75$, the system reuses the stored modality and immediately activates the corresponding control mapping, so the user simply picks the object up and starts using it. If no existing TUI passes the threshold, the system enters an interaction demonstration window: it auto-starts a $3$-second hand-motion capture, returns the top-3 candidate input modalities, and stores the user's selection together with the hand-object view. 

\subsubsection{Interaction Tracking and State Machine}
\label{sec:tracking_sm}

\begin{figure}[t]
\centering
\Description[State machine diagram with states for Idle, Engaged, Touch, Hover and Release interaction phases.]{Five labeled states are connected by arrows. Idle transitions to Engaged through hand-object engagement and input-modality inference; activation then transitions Engaged to Touch. Lifting the hand from the surface changes Touch to Hover, and re-touching returns Hover to Touch. Leaving the object or interaction changes Engaged, Touch, or Hover to Release. A reset arrow returns Release to Idle. Solid arrows show engagement, activation and touch-hover transitions; dotted arrows show leaving and reset transitions.}
  \includegraphics[width=1\columnwidth]{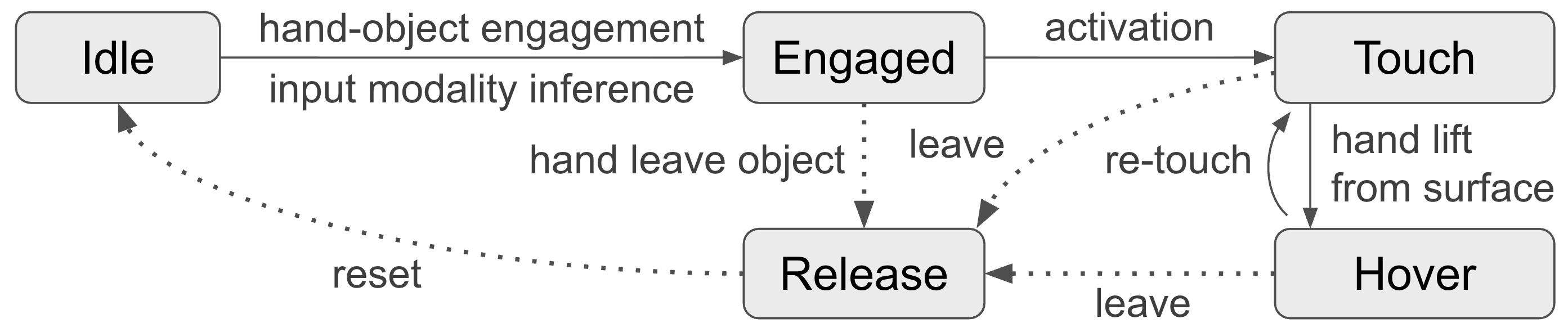}
\vspace{-2mm}
\caption{Midas TUI interaction state model.\vspace{-4mm}}

  \label{fig:tracking-state-machine}
\end{figure}

Once an object is activated, MidasTUI must decide moment-by-moment whether the ongoing hand-object behavior should initiate, maintain, pause, or terminate control. We adopt the event-driven structure used in touch and pointer input systems~\cite{android2024input}, but operate over hand and finger motion observed by the headset cameras and hand tracker rather than over capacitive contacts~\cite{wolf2011taxonomy, sharma2021solofinger, sharma2022design}. 
Figure~\ref{fig:tracking-state-machine} illustrates the shared interaction state structure used across input modalities. While specific thresholds and transitions vary by modality, they follow a common set of interaction phases. 
For each modality, the system monitors a small set of motion cues (surface displacement, angular change, palm vertical velocity, and per-finger vertical velocity) and updates the interaction state accordingly. Following standard pointer-event vocabulary, our state machine uses five states: \emph{Idle}, \emph{Engaged}, \emph{Touch}, \emph{Hover}, and \emph{Release}.

\paragraph{Interaction states.}

\emph{Idle}: the object rests on the surface with no engagement and no control output.
\emph{Engaged}: hand-object engagement is confirmed and a modality is bound, but the system is still observing whether motion is sufficiently stable and intentional to start emitting control events.
\emph{Touch}: the modality-specific activation condition is met and the system continuously emits control output or fires discrete events. Linear primitives (slider, mouse) activate on $\geq 5$\,mm of on-surface displacement; rotary primitives (knob) activate on $\geq 5^{\circ}$ of angular change. Discrete clicks are detected from the index fingertip's smoothed vertical velocity: the finger must first cross a downward threshold of $-0.05$\,m/s, then an upward threshold of $+0.02$\,m/s, with a minimum vertical travel of $8$\,mm. To suppress false clicks during whole-hand lifting, the wrist must remain below $0.08$\,m/s during the click, and successive clicks are separated by a $300$\,ms cooldown.
\emph{Hover}: the user remains engaged with the object but lifts and repositions it before re-contacting the surface. As with mouse clutching during relative control, this lets users reset the workspace without affecting output. Hover is engaged when the smoothed palm vertical velocity exceeds $0.05$\,m/s upward. While in Hover, cursor movement stops, slider values freeze and pen strokes terminate; on landing back into Touch, sliders resume from the held value and a new pen stroke begins.
\emph{Release}: the system aborts the current interaction when the hand leaves the object, when the engagement signal becomes unreliable, when motion no longer matches the inferred modality, or when tracking is lost for an extended period. After Release, the system resets to \emph{Idle}.

\begin{figure*}[t]
  \centering
  \Description[Six AR application scenarios showing everyday objects mapped to interface controls in a Photoshop workspace.]{Six panels showing MidasTUI application scenarios in an AR Photoshop workspace: (a) user scenario overview, (b) a book cover used for button clicks, (c) a key used for slider adjustment, (d) a coffee cup used as a knob for image rotation, (e) an earbud case used as a mouse for cursor navigation, (f) an eyedropper used for freehand drawing as a pen.}
  \includegraphics[width=1\linewidth]{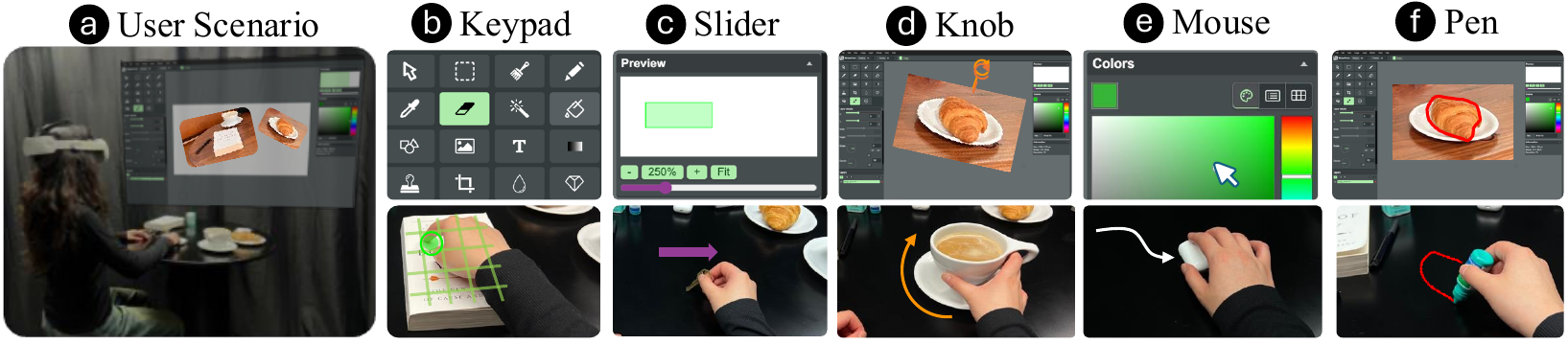}
  \vspace{-4mm}
  \caption{MidasTUI enables diverse interactions using available daily objects within a web-based Photoshop workspace. Examples include: (a) User scenario. (b) Book cover for button clicks. (c) Key for slider adjustment. (d)  Coffee cup for image rotation. (e) Earbud case for cursor navigation. (f) Eyedropper for freehand drawing to crop an image.}
  \label{fig:application-scenarios}
  \vspace{-3mm}
\end{figure*}


\subsection{Interaction-to-Application Mapping}
\label{sec:mapping}

While the tracking stage decides \emph{when} hand-object motion is valid input, the mapping stage decides \emph{how} that input is interpreted. MidasTUI does not bind objects to specific UI widgets; instead, it binds them to interaction primitives organized along four dimensions, and applications consume those primitives. This indirection lets the same physical interaction generalize across applications that share common control structures, including graphics editing, document review, note taking, and web browsing.

\subsubsection{Interaction Primitive Design Space}
We characterize interaction primitives along four dimensions: temporal form, motion primitive, interaction dimensionality, and control mapping.

\paragraph{Temporal form.}
An interaction may be \emph{continuous} or \emph{discrete}. Continuous interactions emit a stream of values while active, such as cursor movement, parameter adjustment, or drawing. Discrete interactions trigger isolated events, such as a click, tool switch, or mode toggle.

\paragraph{Motion primitive.}
Motion may be primarily \emph{linear} or \emph{rotary}. Linear primitives are driven by translation along one or more axes on the surface (slider, mouse), while rotary primitives use angular change around a dominant axis (knob).

\paragraph{Interaction dimensionality.}
Primitives may be \emph{1D} (scalar adjustment such as a slider), \emph{2D} (planar position, such as cursor or pen), or \emph{3D} (spatial input, such as free-space pointing).

\paragraph{Control mapping.}
Mapping may be \emph{absolute} or \emph{relative}. Absolute mappings tie physical position directly to an input location or discrete region, as in the pen, slider, and keypad control. Relative mappings translate physical changes into incremental updates, as in the knob and mouse control. Relative mappings are useful when the virtual control range exceeds the comfortable physical workspace.

\subsubsection{Representative Interaction Primitives}
\label{interaction_primitives}

Productivity and creative interfaces are built around a small recurring set of controls. Figure~\ref{fig:application-scenarios} illustrates how MidasTUI realizes five of them in a photo-editing application on the wearable display.

\paragraph{Keypad.}
Keypad is a discrete, absolute 2D control for commands or mode selection (Figure~\ref{fig:application-scenarios}b). Objects with a planar surface, such as a book cover, are mapped to an $n{\times}n$ grid of virtual keys. Participants tapped different regions of the physical surface to select the corresponding virtual keys. The selected key is determined by the 2D position of the pointing fingertip on the surface, while a downward fingertip motion, detected from its vertical velocity, triggers the selection.

\paragraph{Slider.}
The slider primitive is a continuous, absolute 1D control for scalar parameters such as size, hue, or zoom (Figure~\ref{fig:application-scenarios}c). Participants translated the engaged physical object along the slider axis on the tabletop. The position of the manipulating hand's palm center along this axis is mapped directly to the corresponding value within the slider's range. The value remains unchanged while the object stays stationary or is lifted from the surface.

\paragraph{Knob.}
Knob is a continuous, relative 1D rotary control for parameters such as element orientation and opacity (Figure~\ref{fig:application-scenarios}d). Participants rotated the engaged physical object on the tabletop to adjust the controlled parameter. The rotation angle is inferred from the angle between the thumb and index fingertips, and changes in this angle are mapped to relative changes in the parameter. Lifting the hand and re-contacting the object clutches the control, allowing incremental rotary adjustments beyond the comfortable range of a single motion.

\paragraph{Mouse.}
Mouse is a continuous, relative 2D control with control-display gain (Figure~\ref{fig:application-scenarios}e). Participants moved the engaged physical object across the tabletop to move the cursor. Cursor movement is derived from changes in the 2D position of the manipulating hand's palm center. Lifting the object and re-contacting the surface clutches the control, allowing extended cursor travel within a small physical area. An index-finger click on the engaged object triggers a selection, and a double-click toggles drag.

\paragraph{Pen.}
Pen is a continuous, absolute 2D control for drawing, tracing, handwriting, and highlighting (Figure~\ref{fig:application-scenarios}f). Participants moved the engaged object across the tabletop to draw. The drawing point on the application window is determined by an absolute spatial mapping from the 2D position of the manipulating hand's palm center on the tabletop to the corresponding 2D position in the application window. Surface contact determines stroke state: lifting the object ends the current stroke, and the next contact begins a new one.

\paragraph{Selecting among same-primitive widgets.}
A single physical object can be reused to control multiple UI widgets that share the same interaction primitive. For example, several sliders may control opacity, hue, and brush size. In such cases, the non-dominant hand uses mid-air pinch-and-swipe gestures to switch among candidate widgets of the same function. The selected widget then receives the input generated by the physical object. For example, the non-dominant hand swipes up and down to switch between the four sliders shown on the left side of the application window (Figure~\ref{fig:application-scenarios}a).

\subsubsection{Bimanual Coordination}
\label{sec:bimanual}

Productivity and creative tasks rarely map onto a single tangible control. Users often need to adjust a value while switching widgets or alternate between tool selection and content manipulation. MidasTUI supports such workflows with a bimanual mode that runs two \emph{Move Together} predictors in parallel, allowing each hand to control a separate TUI.
While both hands can enable any of the five supported controls, MidasTUI also affords asymmetric configurations grounded in classical bimanual division of labor~\cite{guiard1987asymmetric}. 
In one configuration, the dominant hand performs fine-grained continuous control such as a mouse or pen, while the non-dominant hand handles discrete navigation or target selection on a keypad. In another, the non-dominant hand stabilizes a held object as a spatial frame of reference while the dominant hand performs precise adjustments relative to it. These asymmetric pairings let users execute compound tasks in a single coordinated motion rather than serializing them across modes.

\begin{figure}[t]
  \centering
  \Description[Grid of 25 everyday objects used for Move Together data collection and a passthrough camera view.]{Left: a grid of 25 everyday objects used in the Move Together data collection, including mugs, bottles, pens, keys, rings, and electronic accessories. Five small objects were removed due to low detection accuracy. Right: a passthrough camera view from a Meta Quest headset showing a hand interacting with an object on a desk surface.}
  \includegraphics[width=1\columnwidth]{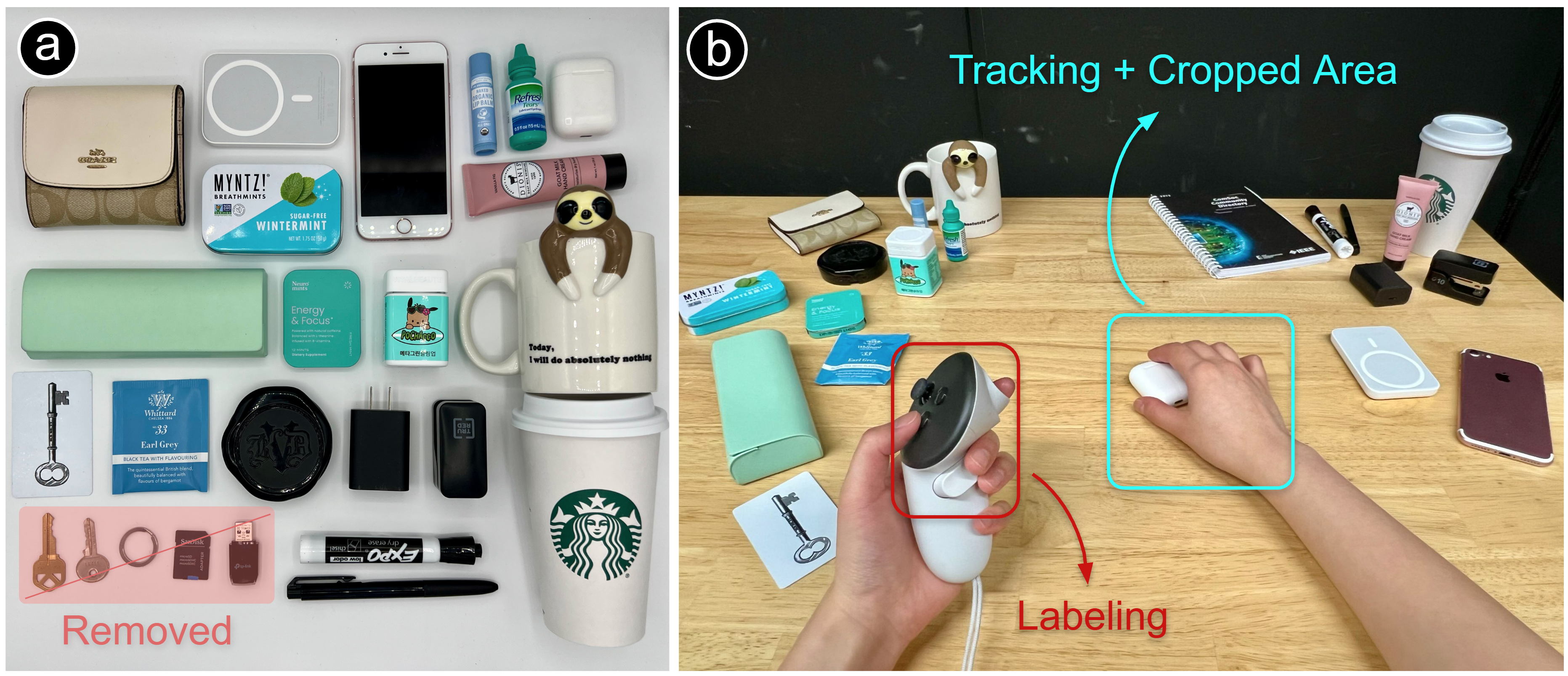}
  \vspace{-2mm}
  \caption{(a) Move together data collected with 25 objects; 5 small objects were removed due to low accuracy. (b) The passthrough camera view of the data collection experiment.}

  \label{fig:move-together-objects}
  \vspace{-4mm}
\end{figure}

\section{MidasTUI Performance Evaluation}
\label{sec:eval}

We conducted two studies to evaluate the performance of two core recognition models in MidasTUI: the hand-object touch and motion detection model, \emph{Move Together}, and the input modality inference model, \emph{Summon}. The study protocol was approved by the institutional review board of the authors' research institution.

\subsection{Hand-Object Engagement Evaluation}
\label{sec:eval-move-together}

\subsubsection{Participants.}
We recruited 16 participants (7 female, 9 male) from a university community (ages 20-33; $M=25$, $SD=4$), all of whom were right-handed. Three participants used VR weekly and two monthly, and the remaining 11 rarely or never used VR. The most commonly reported uses were gaming and research studies. Each session lasted approximately 60 minutes, and participants received USD 20 in compensation.

\subsubsection{Data Collection.}
For Move Together, we collected training and evaluation data using the passthrough cameras of a Meta Quest 3 headset. All hand tracking, hand mask generation, and data collection were implemented directly on the passthrough camera view using the Meta XR All-in-One SDK in Unity on the Meta Quest 3. Participants interacted with 20 everyday objects spanning a range of shapes, sizes, and materials, including mugs, power banks, and etc. (Figure~\ref{fig:move-together-objects}).

For each object, we recorded three interaction sequences following the pattern approach $\to$ grasp $\to$ move $\to$ release $\to$ leave, yielding approximately 3,000 frames per object at 30 FPS. Each frame was annotated with a binary Move Together label. Positive samples included object translation, rotation, and lifting in 3D space, while negative samples included hovering, reaching, and resting contact without displacing the object. The dataset was balanced across positive and negative labels.

We collected the training data from one author and applied augmentation to diversify background appearance and hand color. Despite this limited training source, the model generalized well across all participants in evaluation.

\subsubsection{Temporal Head Comparison.}
We compared the four temporal heads described in Section~\ref{sec:design-pick} on the full 20-object dataset. All models used the same FastViT-T8~\cite{vasu2023fastvit} backbone and were trained end to end with AdamW, cosine learning rate scheduling, and early stopping based on validation F1 (patience = 10). Table~\ref{tab:head-results} summarizes performance across temporal window sizes $K \in \{8,16,32\}$.

\begin{table}[t]
  \centering
    \caption{Hand--object engagement detection (Move~Together)
      results across temporal heads and window sizes $K$.
      Best per column in \textbf{bold}.}
      \vspace{-1mm}
      \Description[Classification results for four temporal heads across window sizes K=8, 16, and 32.]{A table with 12 rows and 6 columns comparing Move Together classification accuracy, F1, AUROC, and parameter count across four temporal heads (Mean, TCN, BiLSTM, Transformer) at window sizes K=8, 16, and 32. The Transformer at K=8 achieves the best accuracy (0.921) and F1 (0.918). The Mean baseline is substantially lower at 0.87 F1.}
    \label{tab:head-results}
    \small
    \tagpdfsetup{table/header-columns={1}}
    \begin{tabular}{llcccc}
      \toprule
      \textbf{Head} & \textbf{$K$} & \textbf{Acc.} & \textbf{F1} & \textbf{AUROC} & \textbf{Params} \\
      \midrule
      Mean & 8   & 0.873 & 0.872 & 0.946 & 3.45M \\
      Mean & 16  & 0.821 & 0.824 & 0.905 & 3.45M \\
      Mean & 32  & 0.781 & 0.788 & 0.886 & 3.45M \\
      \midrule
      TCN & 8   & 0.905 & 0.905 & 0.962 & 3.52M \\
      TCN & 16  & 0.902 & 0.899 & 0.959 & 3.52M \\
      TCN & 32  & 0.904 & 0.902 & 0.960 & 3.52M \\
      \midrule
      BiLSTM & 8   & 0.903 & 0.903 & 0.962 & 6.09M \\
      BiLSTM & 16  & 0.906 & 0.903 & 0.964 & 6.09M \\
      BiLSTM & 32  & 0.906 & 0.906 & 0.964 & 6.09M \\
      \midrule
      Transformer & 8   & \textbf{0.921} & \textbf{0.918} & \textbf{0.968} & 5.06M \\
      Transformer & 16  & 0.900 & 0.899 & 0.959 & 5.06M \\
      Transformer & 32  & 0.898 & 0.896 & 0.957 & 5.06M \\
      \bottomrule
    \end{tabular}
\end{table}

\begin{table}[t]
  \centering
    \small
    \caption{Inference efficiency per temporal head ($K{=}8$)}
    \vspace{-1mm}
    \Description[Inference efficiency per temporal head at K=8 showing parameters, GPU latency, FPS, and deployment FPS.]{A table with 5 rows and 5 columns reporting inference efficiency. GPU processing times range from approximately 5.2 to 5.8 milliseconds per clip on an RTX A6000. The Mean head is fastest at 192 FPS, while the BiLSTM is slowest at 171 FPS. Deployment FPS ranges from 35 to 48 with a Meta Quest 3 headset paired with a laptop. The backbone alone requires 5.0 milliseconds.}
    \label{tab:efficiency}
    \tagpdfsetup{table/header-columns={1}}
    \begin{tabular}{lcccc}
      \toprule
      \textbf{Head} & \textbf{Params} & \textbf{GPU (ms)} & \textbf{FPS} & \textbf{Deployment FPS} \\
      \midrule
      Mean        & 3.45M & 5.2 & 192 & 48 \\
      TCN         & 3.52M & 5.4 & 185 & 45 \\
      BiLSTM      & 6.09M & 5.8 & 171 & 35 \\
      Transformer & 5.06M & 5.6 & 178 & 42 \\
      \midrule
      \textit{Backbone only} & \textit{3.26M} & \textit{5.0} & --- & --- \\
      \bottomrule
    \end{tabular}

\end{table}

\begin{table}[t]
  \centering
  \small

    \caption{Feature ablation for the \emph{generalised} Summon model
      (first four rows), evaluated via LOPO cross-validation on
      16~participants without target-user data at training time.
      The final row reports the \emph{personalised} model,
      adapted using target-user demonstrations.}
    \vspace{-1mm}
    \Description[Feature ablation showing accuracy improvement from position to velocity to acceleration to Mixup.]{A table whose first four rows show feature and regularization ablations for the generalized Summon model under LOPO evaluation, with a final row reporting a personalized model adapted using target-user demonstrations. Position only achieves 0.722 accuracy. Adding velocity improves by 4.7 percentage points to 0.769. Mixup adds 2.5 percentage points to 0.794. Acceleration adds 1.5 percentage points to 0.809. Per-user personalization reaches 0.975.}
    \label{tab:summon-features}
    \tagpdfsetup{table/header-columns={1}}
    \begin{tabular}{lccc}
      \toprule
      \textbf{Features} & \textbf{Dim} & \textbf{Top-1 Acc.} & \textbf{$\Delta$} \\
      \midrule
      Position only              & 78  & 0.722 & ---     \\
      + Velocity                 & 156 & 0.769 & +4.7\,pp \\
      + Mixup ($\alpha{=}0.2$)   & 156 & 0.794 & +2.5\,pp \\
      + Acceleration             & 234 & \textbf{0.809} & +1.5\,pp \\
      \midrule
      \textit{+ Per-user data (personalised)} & \textit{234} & \textit{\textbf{0.975}} & \textit{+16.6\,pp} \\
      \bottomrule
    \end{tabular}
\end{table}

BiLSTM, Transformer, and TCN all achieve F1\,$\approx$\,0.90
at $K{=}8$, while the Mean baseline peaks at 0.87---confirming
that temporal ordering is critical for engagement detection.
At $K{=}8$, the Transformer achieves the highest F1 (0.918),
followed by TCN (0.905) and BiLSTM (0.903).
Increasing $K$ from 8 to 16 yields marginal gains for BiLSTM
(+0.3\,pp accuracy) but degrades the Mean head ($-$5.2\,pp)
and TCN ($-$0.3\,pp). Longer windows do not consistently
improve performance across heads, and the Transformer achieves
its best performance at $K{=}8$ among the tested window sizes.

\subsubsection{Inference Efficiency.}
Table~\ref{tab:efficiency} reports inference latency at $K=8$ on an NVIDIA RTX A6000, as well as runtime in our deployment setup using a Meta Quest 3 headset paired with a laptop equipped with an Intel Core i7 processor and an NVIDIA RTX 3080 GPU.

GPU processing times range from approximately 5.2 to 5.8 ms per clip across all temporal heads on the RTX A6000, indicating that the temporal module introduces only limited overhead beyond the FastViT backbone. Since the backbone alone requires 5.0 ms, most of the computation is dominated by visual feature extraction rather than temporal modeling. The Mean and TCN heads are the most efficient, reaching 192 and 185 FPS, respectively, while BiLSTM is the slowest at 171 FPS. The Transformer achieves 178 FPS on the RTX A6000 and 42 FPS in our deployment setup, which remains sufficient for real-time use.

\begin{figure}[t]
  \centering
  \Description[Two line plots showing F1 score and inference latency versus window size K for four temporal heads.]{Both plots compare window sizes K=8, 16, and 32. Circles identify BiLSTM, squares TCN, triangles Transformer, and diamonds Mean pooling. Left: Transformer has the highest F1, approximately 0.92 at K=8; BiLSTM and TCN remain near 0.90, while Mean pooling falls from approximately 0.87 to 0.79 as K increases. Right: all heads take approximately 5 to 6 milliseconds at K=8. At K=32, BiLSTM takes approximately 14 milliseconds, TCN and Transformer approximately 13 milliseconds, and Mean pooling approximately 7 milliseconds.}
  \includegraphics[width=1\columnwidth]{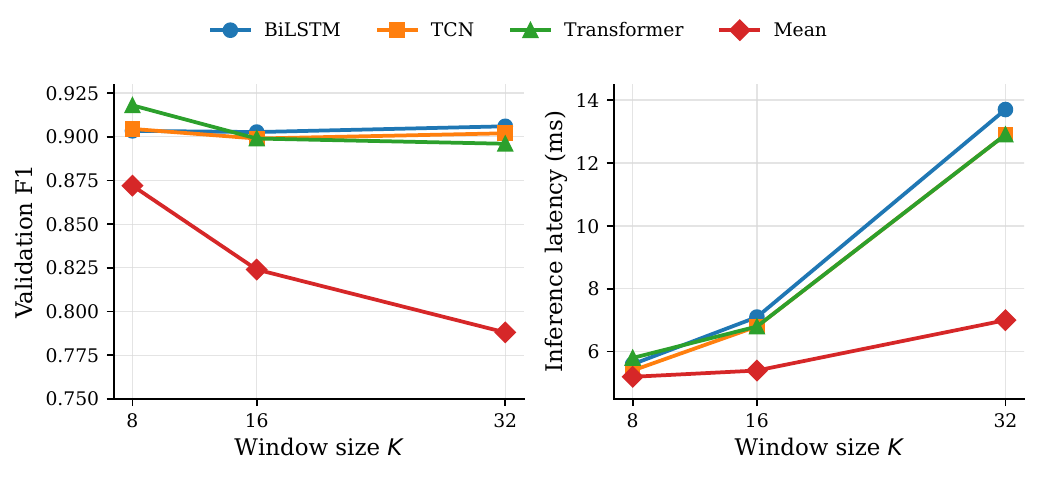}
  \caption{Validation F1 and inference latency versus window size $K$. The Transformer achieves the best F1 (0.92 at $K=8$). Latency increases with $K$.\vspace{-3mm}}

  \label{fig:f1-latency-vs-k}
\end{figure}

Considering both recognition accuracy and inference cost (Figure~\ref{fig:f1-latency-vs-k}), we select the Transformer head for the final Move Together model. Although it is not the fastest variant, it achieves the best overall performance at $K=8$ (Table~\ref{tab:head-results}), while still satisfying the real-time constraints of our headset plus laptop system.

\subsection{Input Modality Inference Evaluation}
\label{sec:eval-summon}

The Summon (input modality inference) stage classifies the user's
demonstrated hand motion into one of ten input device classes (Figure~\ref{fig:summon-objects}). The same group of participants who completed the move-together evaluation also participated in the summon evaluation.

\subsubsection{Data Collection.}
We collected Summon training and evaluation data from 16 participants using the passthrough cameras of a Meta Quest 3 headset. All hand tracking, hand mask generation, and data collection were implemented directly on the passthrough camera view using the Meta XR All-in-One SDK in Unity on the Meta Quest 3. Participants performed demonstrations with 10 interaction modalities: slider, knob, switch, mouse, touchpad, pen, joystick, d-pad, keypad, and pointer.
Each participant performed four demonstrations for each modality, yielding 40 sequences per participant and 640 sequences in total. The headset’s built-in hand tracking recorded the 3D world-space positions of 26 hand joints at each timestep.

\subsubsection{Feature Ablation.}
We augment raw joint positions with temporal derivatives
and evaluate each feature set via leave-one-participant-out
(LOPO) cross-validation (Table~\ref{tab:summon-features}).

Velocity provides the largest single gain (+4.7\,pp),
capturing motion dynamics that static positions cannot encode.
Mixup regularization~\cite{zhang2017mixup} contributes +2.5\,pp
by smoothing decision boundaries across the small per-class
sample size.
Acceleration adds a further +1.5\,pp by distinguishing
impulsive gestures (e.g., keypad tap) from sustained motion
(e.g., slider drag).
The generalized model achieves \textbf{80.9\%} top-1 and
\textbf{90.8\%} top-3 accuracy across 16~unseen participants,
without requiring any target-user data at training time.

\subsubsection{Per-Class Performance.}
Table~\ref{tab:summon-perclass} reports per-class recall for
both the generalized model~(\ref{tab:summon-gen}) and the
personalized model~(\ref{tab:summon-pers}). 
In the generalized model (Table~\ref{tab:summon-gen}), easy classes (\emph{pen}, \emph{mouse}, \emph{pointer}) achieve top-1 recall ${\geq}$0.91, while hard classes (\emph{joystick}, \emph{switch}) fall below 0.69.
Top-3 recall recovers 10\,pp on average; for \emph{joystick}, the gain is 22\,pp (0.64\,$\to$\,0.86), suggesting that the correct class is often ranked 2nd or 3rd. The personalized model (Table~\ref{tab:summon-pers}) achieves perfect recall on 8 of 10 classes, with only slider (0.88) and knob (0.88) producing any errors.

\begin{table}[t]
  \centering
  \begin{minipage}[t]{\linewidth}
    \small
    \setlength{\tabcolsep}{3pt}
    \vspace{0mm}
    \centering
    \caption{Per-class recall for Summon classification.
      \textbf{(a)}~Generalized model (LOPO).
      \textbf{(b)}~Personalized model.\vspace{-3mm}}
      \Description[Side-by-side per-class recall tables for generalized and personalized Summon models.]{Two subtables showing per-class top-1 and top-3 recall. Left (generalized, blue-tinted): overall top-1 is 0.81 and top-3 is 0.91. Pen achieves 0.97 top-1 while joystick is lowest at 0.64. Right (personalized, green-tinted): overall top-1 is 0.98 and top-3 is 0.99. Eight of ten classes achieve perfect recall, with only slider and knob at 0.88.}
    \label{tab:summon-perclass}
    \vspace{1mm}
    \renewcommand{\arraystretch}{1.5}
    \begin{subtable}[t]{0.48\linewidth}
      \centering
      \phantomsubcaption
      \textbf{(a) Generalized (LOPO)}\par
      \label{tab:summon-gen}
      \tagpdfsetup{table/header-columns={1}}
      \begin{tabular}{lcc}
        \toprule
        \textbf{Class} & \textbf{Top-1} & \textbf{Top-3} \\
        \midrule
        \rowcolor{blue!6}
        pen       & 0.97 & 1.00 \\
        \rowcolor{blue!6}
        mouse     & 0.92 & 1.00 \\
        \rowcolor{blue!6}
        pointer   & 0.91 & 0.94 \\
        \rowcolor{blue!6}
        touchpad  & 0.86 & 0.94 \\
        \rowcolor{blue!6}
        dpad      & 0.83 & 0.95 \\
        \rowcolor{blue!6}
        keypad       & 0.77 & 0.88 \\
        \rowcolor{blue!6}
        knob      & 0.77 & 0.89 \\
        \rowcolor{blue!6}
        slider    & 0.75 & 0.86 \\
        \rowcolor{blue!6}
        switch    & 0.69 & 0.77 \\
        \rowcolor{blue!6}
        joystick  & 0.64 & 0.86 \\
        \midrule
        \rowcolor{blue!12}
        \textit{Overall} & \textit{0.81} & \textit{0.91} \\
        \bottomrule
      \end{tabular}
    \end{subtable}
    \hfill
    \begin{subtable}[t]{0.48\linewidth}
      \centering
      \phantomsubcaption
      \textbf{(b) Personalized}\par
      \label{tab:summon-pers}
      \tagpdfsetup{table/header-columns={1}}
      \begin{tabular}{lcc}
        \toprule
        \textbf{Class} & \textbf{Top-1} & \textbf{Top-3} \\
        \midrule
        \rowcolor{green!8}
        pen       & 1.00 & 1.00 \\
        \rowcolor{green!8}
        mouse     & 1.00 & 1.00 \\
        \rowcolor{green!8}
        pointer   & 1.00 & 1.00 \\
        \rowcolor{green!8}
        touchpad  & 1.00 & 1.00 \\
        \rowcolor{green!8}
        dpad      & 1.00 & 1.00 \\
        \rowcolor{green!8}
        keypad       & 1.00 & 1.00 \\
        \rowcolor{green!8}
        switch    & 1.00 & 1.00 \\
        \rowcolor{green!8}
        joystick  & 1.00 & 1.00 \\
        \rowcolor{green!8}
        slider    & 0.88 & 1.00 \\
        \rowcolor{green!8}
        knob      & 0.88 & 0.88 \\
        \midrule
        \rowcolor{green!15}
        \textit{Overall} & \textit{0.98} & \textit{0.99} \\
        \bottomrule
      \end{tabular}
    \end{subtable}
  \end{minipage}
\end{table}

\begin{figure}[t]
  \centering
    \Description[Ten-by-ten confusion matrix heatmap for Summon classification across 16 LOPO folds.]{A blue-shaded confusion matrix heatmap with 10 rows (true classes) and 10 columns (predicted classes). Diagonal cells show per-class recall as counts and percentages. The strongest diagonal entries are pen (97 percent), mouse (92 percent), and pointer (91 percent). Red-bordered off-diagonal cells highlight the largest confusions: joystick predicted as switch (14 errors), switch predicted as joystick (9 errors), and keypad predicted as touchpad (10 errors).}
  \includegraphics[width=1\linewidth]{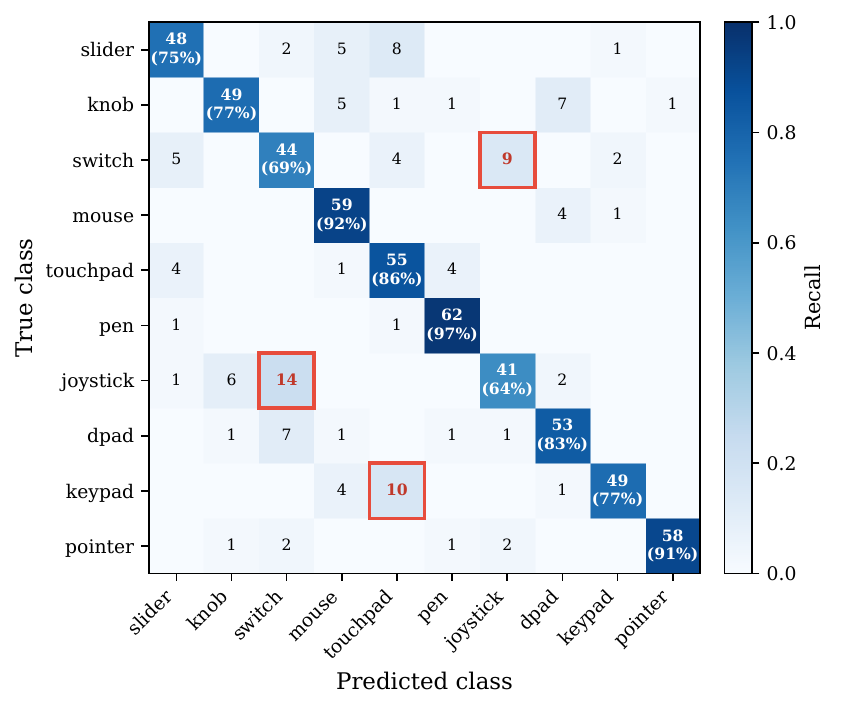}
    \caption{Confusion matrix for Summon classification across 16 LOPO folds. Diagonal entries show per-class recall, and red borders highlight the largest confusion patterns.\vspace{-3mm}}

    \label{fig:summon-confusion}
\end{figure}

\subsubsection{Confusion Analysis.}
Figure~\ref{fig:summon-confusion} shows the pooled confusion matrix. The most prominent error pattern is between \emph{joystick} and \emph{switch}: 14 joystick samples are misclassified as switch, and 9 switch samples are misclassified as joystick, indicating that these two modalities share similar short directional hand motions. Another notable confusion occurs from \emph{keypad} to \emph{touchpad}, with 10 keypad samples predicted as touchpad, likely because both interactions involve a brief downward finger movement.

\subsubsection{Personalized Model.}
In deployment, MidasTUI begins with the generalized model, which presents the user with the top 3 most likely interaction modalities immediately after an object demonstration.
As the user selects from these suggestions over time, the system collects labeled demonstrations that reflect that individual’s motion style and interaction preferences.
The personalized model achieves \textbf{97.5\%} top-1 accuracy
(Table~\ref{tab:summon-pers}), a +16.6\,pp gain over the
generalized model, suggesting that the remaining LOPO errors
stem from cross-user motion variability, which a small amount
of per-user data effectively resolves.

\section{MidasTUI Comparative Exploratory Study}
We conducted a user study to evaluate MidasTUI in a photo-collage task in MR. We compared MidasTUI with mid-air hand interaction to investigate whether, when users interact with a wearable display without a dedicated input device, the tangible interfaces enabled by MidasTUI using everyday objects provide advantages over performing the same interactions entirely in mid-air. Participants completed a photo-collage involving selection, tool switching, object manipulation, and parameter adjustment. We further examined one- and two-handed interaction to understand whether tangible interfaces can support and benefit from bimanual manipulations.

\subsection{Tasks}

\begin{figure*}[t]
    \centering
    \Description[
    Photo-collage task showing the initial collage, five editing stages, and final collage.
    ]{
    The figure illustrates the photo-collage task used in the study. The initial collage is shown on the left, followed by five editing stages in the center: selecting and dragging visual elements, resizing elements by adjusting width and height, rotating elements by changing their angle, adjusting element opacity, and drawing using a color palette and pen tool. The completed target collage is shown on the right.
    }
  \includegraphics[width=\linewidth]{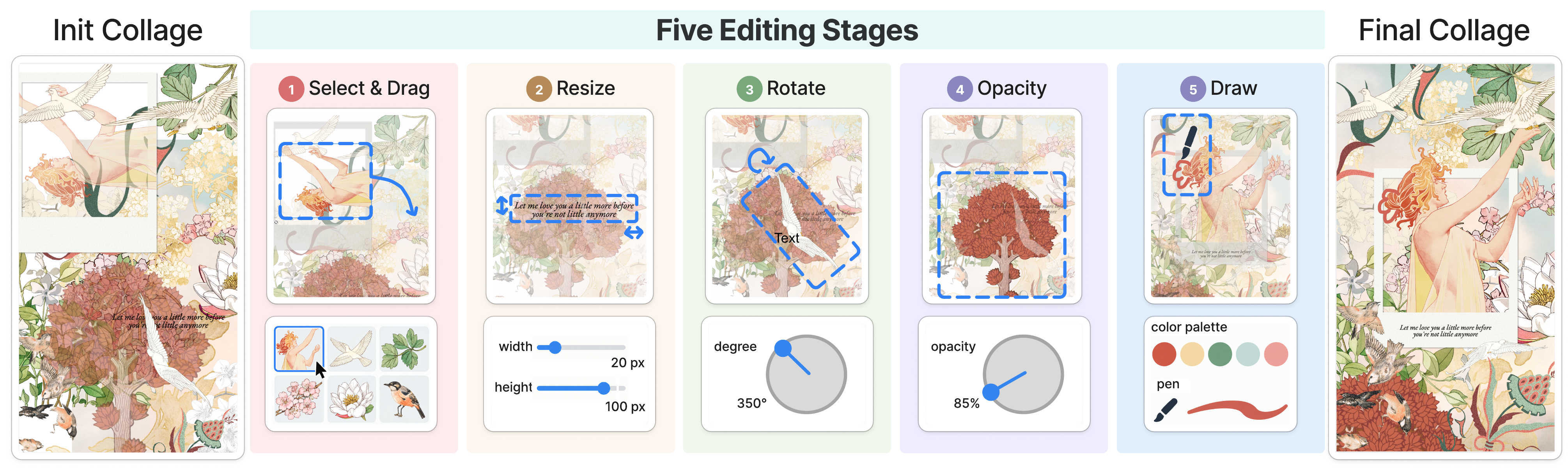}
    \caption{Photo-collage task used in the user study. Participants transformed the initial collage into the target collage through five editing stages: (1) selecting and dragging visual elements, (2) resizing elements, (3) rotating elements, (4) adjusting opacity, and (5) drawing with a selected color and pen tool.}

    \label{fig:study-task}
\end{figure*}

We designed a photo-collage task representative of lightweight work on wearable displays (Figure~\ref{fig:study-task}). Participants were shown a reference collage and asked to reproduce it from a collection of visual elements on a 2D virtual window. The task consisted of five main editing stages. Participants first selected and dragged image elements to their target locations, changed their size by parameter control, then rotated selected elements to match the target orientation, adjusted their opacity to reproduce the desired visual composition, and finally created a color palette and added a small amount of freehand drawing to the collage.

These operations required the five application controls supported by our prototype: \emph{cursor}, \emph{keypad}, \emph{slider}, \emph{dial}, and \emph{pen}. Cursor input was used for selecting and repositioning visual elements. The keypad supported discrete input such as tool selection, color picker, and layer selection. The dial controlled rotation and opacity while the slider controlled parameters such as size and color hues. The pen supported freehand drawing. The task therefore required participants to switch between discrete and continuous input as well as between coarse positioning and fine parameter adjustment.

\subsection{Interaction Techniques}
We compared two hand interaction techniques: \emph{Mid-Air} and \emph{MidasTUI}. Mid-Air served as the baseline because both techniques are controller-free and rely on hand tracking, while MidasTUI additionally grounds interaction in everyday physical objects.

\begin{figure}[b]
    \centering
    \Description[
    Two-panel study setup comparing Mid-Air and MidasTUI interaction.
    ]{
    The figure shows the two interaction conditions used in the user study.
    (a) In the Mid-Air condition, the participant interacts with the photo-collage interface using bare-hand gestures in front of the virtual display.
    (b) In the MidasTUI condition, the same photo-collage interface is shown together with everyday objects placed on the tabletop, which are used as tangible controls for interaction.
    }
  \includegraphics[width=1\linewidth]{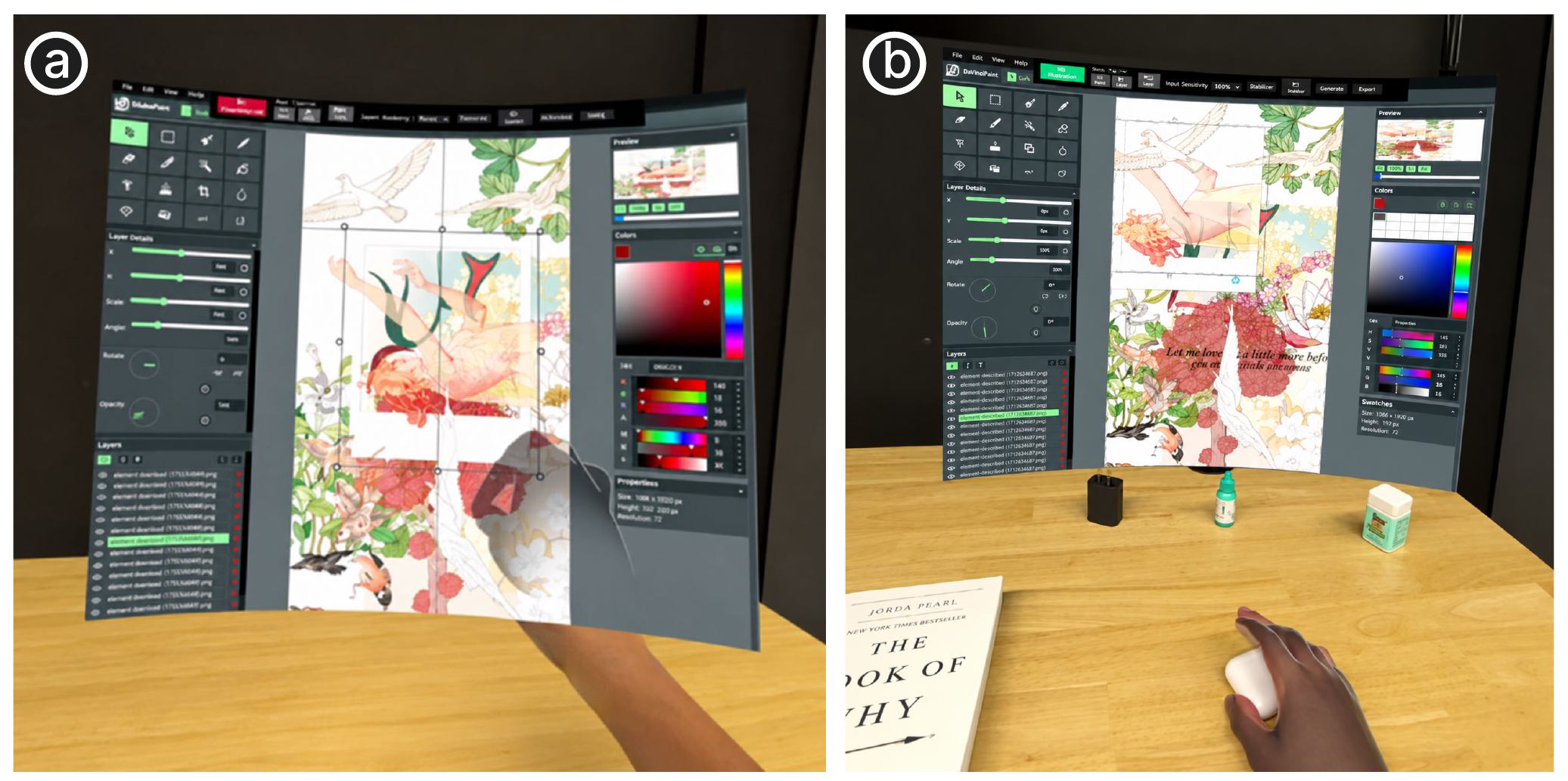}
    \caption{Study setup for the two interaction techniques.
    (a) Mid-Air interaction, where participants manipulate the photo-collage interface using bare-hand gestures.
    (b) MidasTUI interaction, where participants use everyday physical objects as tangible controls for the same photo-collage task.}

    \label{fig:study-setup}
\end{figure}

\subsubsection{Mid-Air Gestures}
In the Mid-Air condition, participants interacted with the application using bare-hand gestures tracked by the headset (Figure~\ref{fig:study-setup}a). Participants controlled a visual ray rendered between the thumb and index finger, which was cast onto the virtual window to control the cursor. To select tools, participants moved the cursor to the target and pinched to select it~\cite{metaHand2024}. The same point-and-pinch interaction was used to activate buttons, manipulate sliders and dials, drag visual elements, and drawing.

We supported both one- and two-handed interaction. In the one-handed condition, participants performed manipulations sequentially using a single cursor controlled by the dominant hand. In the two-handed condition, each hand controlled an independent cursor using the same interaction technique, enabling symmetric bimanual interaction.

\subsubsection{MidasTUI}
In the MidasTUI condition, participants performed the same operations by manipulating everyday objects as tangible proxies for virtual controls. Participants were provided with five everyday objects corresponding to different interaction primitives: (1) an earbuds case for cursor control, (2) a charger adapter for slider input, (3) a pill box for dial input, (4) a book for keypad input, and (5) an eye-drop bottle for pen input (Figure~\ref{fig:study-setup}).

To register an everyday object as a TUI and activate its associated function, participants first brought the object into the interaction area (i.e., the tabletop), grasped it, and briefly demonstrated the intended movement for the target operation for approximately 3--4~s. For example, participants moved the earbuds case across the tabletop to register it for cursor control. Once an object had been registered and assigned a function, it could be reused without demonstration. Translation of an engaged object provided cursor, slider, or pen input; rotation provided dial input; and tapping supported confirmation or discrete keypad input as described in Section~\ref{interaction_primitives}.

MidasTUI also supports both one- and two-handed interactions. In the one-handed condition, participants switched among tangible controls sequentially using the dominant hand. In the two-handed condition, participants could manipulate different physical objects with each hand, enabling asymmetric bimanual interaction. For example, the non-dominant hand could move the cursor to select an object while the dominant hand adjusted its rotation using the dial control.


\subsection{Apparatus}
The study was conducted using a Meta Quest~3 headset with video passthrough and hand tracking. Both MidasTUI and the Mid-Air interaction technique were implemented in Unity using the Meta XR All-in-One SDK and operated on the headset's passthrough camera view. The headset was paired with a laptop equipped with an Intel Core i7 processor and an NVIDIA RTX 3080 GPU. Participants were seated at a desk, with the virtual window positioned in front of them and the everyday objects used as MidasTUI controls placed within comfortable reach on the tabletop. The Mid-Air condition used the same headset, application, and visual workspace, but did not require participants to manipulate physical objects.

\subsection{Participants}
We recruited 10 participants (6 male and 4 female) from a university, ranging in age from 22 to 30 years ($M=25.0$, $SD=2.7$). Participants generally had varying prior experience with wearable displays. Two participants reported weekly use, three reported using them rarely, and the rest of them never using AR/VR or smart-glasses devices. Most participants were familiar with the input devices that were mimicked in this study such as button, slider, mouse, knob, and pen. Seven participants reported daily use, two monthly use, and one reported no regular use. Each study session lasted approximately 50 minutes, and participants received USD 20 in compensation. The study followed the ethics review requirements of the authors' research institution and was approved by its institutional review board.

\subsection{Study Design and Procedure}
We used a within-participant design with \emph{Interaction Technique} (\emph{MidasTUI} vs.\ \emph{Mid-Air}) as the independent variable. Each participant completed the photo-collage task using both techniques, and the order of the two conditions was counterbalanced across participants. Within each technique, participants were encouraged to experience both one- and two-handed interaction freely. For example, they could use the dominant hand to select an element, and use the non-dominant hand to perform rotation.

At the beginning of each session, participants completed a demographic questionnaire and watched a video demonstration accompanied by instructions from the experimenter, which introduced the task and interaction techniques, followed by a short practice period. Participants then completed the task under both interaction-technique conditions. After each condition, they completed questionnaires assessing workload, system usability, and interaction-specific feedback. After completing both techniques, participants directly compared the two interaction techniques through a customized survey and provided open-ended comments on the advantages and limitations of each technique.

\begin{figure*}[t]
  \centering
  \Description[
    Three-panel boxplots comparing NASA-TLX workload, SUS scores, and task
    completion time for MidasTUI and Mid-Air.
    ]{
    Three panels summarize participants' evaluations of MidasTUI and Mid-Air.
    (a) Each pair has Mid-Air on the left and MidasTUI on the right. NASA-TLX workload ratings cover mental demand, physical demand, temporal
    demand, performance, effort, and frustration. MidasTUI shows lower physical
    demand, effort, and frustration than Mid-Air, while perceived performance is
    higher for MidasTUI.
    (b) SUS scores for Mid-Air and MidasTUI, with individual participant scores
    overlaid on the boxplots; MidasTUI receives higher usability ratings overall.
    (c) Task completion time for the two interaction techniques, with individual
    participant times overlaid on the boxplots. Participants generally complete
    the task faster with Mid-Air than with MidasTUI.
    }
  \includegraphics[width=\linewidth]{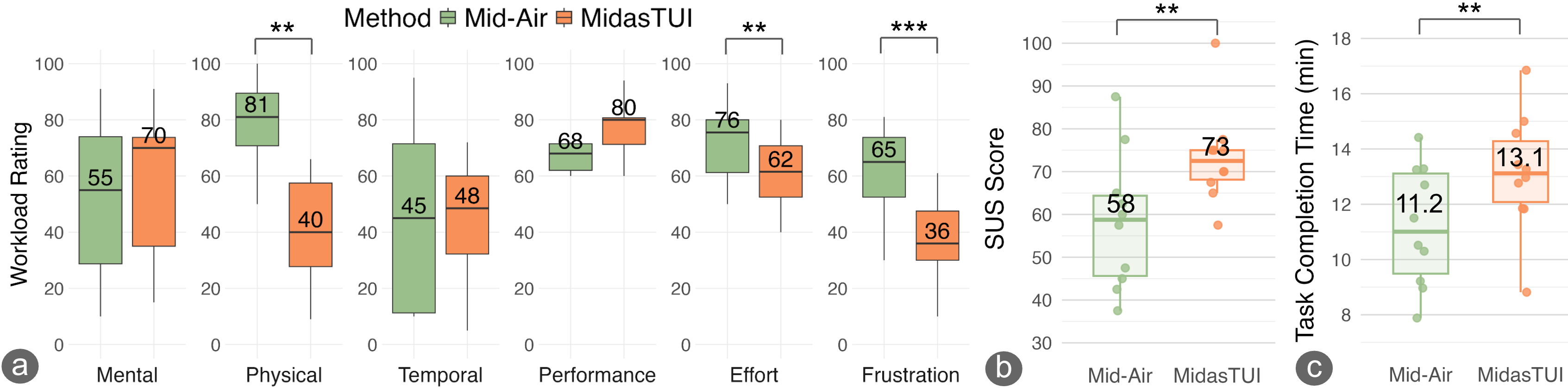}
  \caption{Study measures for (a) NASA-TLX workload, (b) SUS scores, (c) task completion time when interacting with MidasTUI and Mid-Air. In every pair of boxplots, Mid-Air is on the left and MidasTUI on the right.}

  \label{fig:study-workload-sus-ct}
\end{figure*}

\subsection{Evaluation Metrics}
\subsubsection{Workload.}
Workload was measured using the NASA-TLX \cite{hart1988development}, which evaluated mental, physical, and temporal demand, performance, effort, and frustration.

\subsubsection{Usability.}
Participants also completed the System Usability Scale (SUS)~\cite{brooke1996sus} after each interaction condition to assess perceived system usability.

\subsubsection{Task Completion Time.}
We recorded the time required to complete the photo-collage task under each interaction condition, measured from the start of task interaction until the participant completed the final operation.

\subsubsection{Self-reported Feedback.}
We collected interaction-specific feedback after each condition. For MidasTUI, participants rated the ease of switching tools by changing physical objects and switching among controls associated with the same interaction primitive. For Mid-Air, participants rated the ease of navigating to and selecting tools using mid-air gestures. Participants also reported their preferred and least preferred interactions and described the advantages and limitations of each technique in open-ended responses. After completing both conditions, participants completed a customized comparative survey asking which interaction technique they perceived as more natural, precise, stable, and responsive, as well as which technique they preferred overall.

\subsection{Data analysis}
For continuous measures, normality was assessed using the Shapiro--Wilk test. Measures that satisfied the normality assumption were analyzed using repeated-measures ANOVA, followed by pairwise paired $t$-tests with Bonferroni correction when post-hoc comparisons were needed. For continuous measures that did not satisfy the normality assumption, we used the Aligned Rank Transform (ART)~\cite{wobbrock2011aligned}, followed by ART-C~\cite{elkin2021aligned} for post-hoc comparisons when needed. For ordinal measures, we used paired Wilcoxon signed-rank tests with Bonferroni correction. We used a significance level of $p=0.05$ for all tests.

\begin{samepage}

\paragraph{Use of AI Tools.}
The authors used generative AI tools (ChatGPT, Claude, and Codex) to assist with model implementation and the development of portions of the system software; support figure and appendix scripts; check numerical and statistical consistency in reported results; revise selected manuscript passages, figure captions, and table labels; and provide language editing and proofreading. Inputs provided to these tools: manuscript drafts, LaTeX source, and reported figures and tables. The authors reviewed the suggested revisions. Authorship boundaries: the authors retain full responsibility for the implementation, research, and final manuscript; AI tools are not credited as authors.
\par\end{samepage}

\subsection{Results}

\subsubsection{NASA-TLX Workload}
Figure~\ref{fig:study-workload-sus-ct}a presents participants' ratings for each workload factor. Interaction technique had a significant effect on physical demand ($F_{(1,9)}=20.13,\ p<0.01$), effort ($F_{(1,9)}=11.76,\ p<0.01$), and frustration ($F_{(1,9)}=24.98,\ p<0.001$). MidasTUI was rated as less physically demanding ($median=40,\ IQR=29.8$) than Mid-Air ($median=81,\ IQR=18.8$), required less effort ($median=61.5,\ IQR=18.2$) than Mid-Air ($median=75.5,\ IQR=18.8$), and was less frustrating ($median=36,\ IQR=17.5$) than Mid-Air ($median=65,\ IQR=21.2$). No significant effects were found for mental demand ($F_{(1,9)}=0.51,\ p>0.05$), temporal demand ($F_{(1,9)}<0.01,\ p>0.05$), or performance ($F_{(1,9)}=4.23,\ p>0.05$).

\subsubsection{System Usability}
SUS scores differed significantly between the two interaction techniques ($F_{(1,9)}=12.37,\ p<0.01$), as shown in Figure~\ref{fig:study-workload-sus-ct}b. MidasTUI received a higher mean SUS score ($M=73.2$, $SD=11.1$) than Mid-Air ($M=58.2$, $SD=15.9$).

\subsubsection{Task Completion Time}
Task completion time differed significantly between the two interaction techniques ($F_{(1,9)}=11.16,\ p<0.01$), as shown in Figure~\ref{fig:study-workload-sus-ct}c. Participants completed the task faster using Mid-Air ($M=11.20$ min, $SD=2.17$) than MidasTUI ($M=13.14$ min, $SD=2.15$).

\subsubsection{Tool and Control Switching}
\begin{figure*}[t]
  \centering
  \Description[
    Two-panel figure comparing switching ratings and preferences for MidasTUI and Mid-Air.
    ]{Two panels summarize participants' subjective evaluations.
    (a) Stacked Likert bars compare MidasTUI and Mid-Air for two statements: ease of switching tools and ease of switching controls assigned to the same tool. MidasTUI receives more positive responses for switching tools, while ratings for switching controls are more similar between the two techniques.
    (b) Stacked horizontal bars show participants' direct preferences for overall preference, stability, precision, naturalness, and responsiveness. Color distinguishes Mid-Air from MidasTUI, while solid and hatched segments indicate one-handed and two-handed interaction, respectively. Across the five measures, participants generally favor MidasTUI, with two-handed MidasTUI accounting for a substantial portion of the preferences.
    }
  \includegraphics[width=\linewidth]{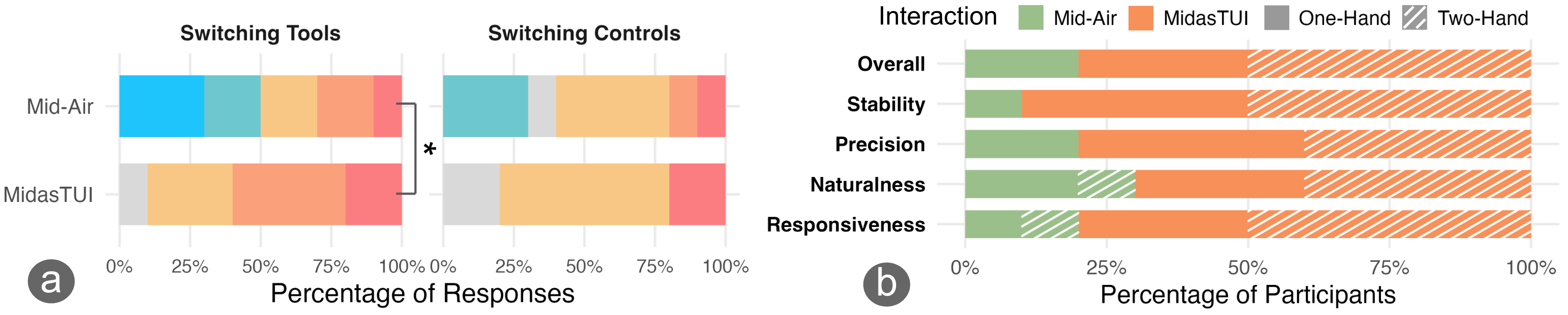}
  \caption{Subjective comparison of interaction techniques. (a) Participants' ratings for switching tools and switching controls using MidasTUI and Mid-Air. (b) Participants' preferences for interaction techniques in terms of naturalness, precision, stability, responsiveness, and overall preference. Mid-Air occupies the left portion of each preference bar and MidasTUI the right; solid and hatched segments indicate one-handed and two-handed interaction, respectively.}

  \label{fig:study-switching-preference}
\end{figure*}

For switching tools (Figure~\ref{fig:study-switching-preference}a), MidasTUI received a higher median rating ($median=6.0,\ IQR=1.0$) than Mid-Air ($median=4.0,\ IQR=3.5$), and the difference is significant ($W=26.5,\ p<0.05$). For switching controls assigned to the same tool, no significant difference was found between the two techniques ($W=15.5,\ p=0.34$). Both received a median rating of 5.0, with a narrower distribution for MidasTUI ($IQR=0$) than for Mid-Air ($IQR=1.75$).

\subsubsection{Preference and User Feedback}
Participants' generally preferred MidasTUI as shown in Figure~\ref{fig:study-switching-preference}b. Eight of ten participants preferred MidasTUI overall, while two preferred Mid-Air. Among the eight participants who preferred MidasTUI, five selected the two-handed configuration and three selected the one-handed configuration; no participant selected Mid-Air Two-Hand as their overall preferred method. MidasTUI was also preferred by 7/10 participants in terms of naturalness, 8/10 for precision, 9/10 for stability, and 8/10 for responsiveness.

Participants who preferred MidasTUI frequently highlighted the physical grounding provided by everyday objects and the resulting sense of control. P3 noted that MidasTUI was easier than interacting in mid-air and that they `liked being able to hold an object.'' P9 similarly appreciated that the tangible controls were `easy to control and find in our life,'' and described their use as feeling natural. Participants also valued bimanual interaction. P5 reported that double-handed interaction `helps a lot to catch the target,'' while P8 described it as `easy to operate and less burden on each hand.'' P4 further valued being able to use both hands and switch among different control methods through gestures.

At the same time, participants identified limitations in the current MidasTUI implementation. Several comments concerned unreliable hand--object contact detection and fine-grained manipulation. P9 reported that the system sometimes continued detecting hand movement after the hand had left an object, leading to unintended responses. P5 described mouse clicking as ``not super controllable and stable,'' while P1 found the knob and slider difficult to use accurately. Bimanual interaction also introduced challenges for some participants: P3 found the non-dominant-hand pinch gesture difficult to trigger reliably, and P9 noted that manipulating objects with both hands could be constrained by the available physical workspace.

Participants valued Mid-Air interaction for its device-free simplicity. P10 described it as `simple to use, all you need is the two hands,'' while P3 appreciated being able to perform the task entirely with hand gestures. However, fatigue and reduced precision were recurring concerns. P9 reported that their `arms are sore after interacting for about 20 mins,'' and P8 similarly noted that the high physical demand caused arm soreness. P4 reported that larger arm movements made the technique more tiring and that Mid-Air controls felt less natural and precise. Fine-grained adjustments were also difficult: P3 specifically reported challenges with rotation, opacity adjustment, and object movement, while P10 found small slider and dial adjustments difficult to control.

Overall, participants' feedback highlights complementary trade-offs between the two techniques. MidasTUI provided physical grounding, tangible feedback, and support for asymmetric bimanual interaction, which participants associated with greater control, stability, and reduced physical burden. Its main limitations arose from hand--object state detection, precise control mappings, and the physical workspace required for tangible manipulation. Mid-Air interaction required no external objects and offered direct access using only the hands, but participants frequently reported greater fatigue and difficulty with fine-grained control.

\section{Discussion}
\label{sec:discussion}
Our evaluations provide evidence at two complementary levels. The technical evaluations show that MidasTUI can reliably detect hand-object engagement and infer demonstrated input modalities across diverse objects and users. Our comparative exploratory user study further evaluates the resulting tangible interaction in practice, showing that grounding input in everyday objects reduces perceived physical burden and improves perceived usability and control relative to Mid-Air interaction, but also increases task completion time. Together, these results suggest that using everyday objects as opportunistic TUIs offers a balance between availability, physical grounding, control, and interaction efficiency. Based on our findings, we first discuss what the technical evaluations reveal about hand-object engagement detection and input modality inference. We then consider the interaction-level benefits and costs of tangible interaction observed in the user study. Finally, we derive broader design implications for opportunistic tangible interaction.

\subsection{Hand--Object Engagement Detection}
\label{sec:disc-engagement}

Two findings from the Move Together evaluation shape how engagement detection should be designed. First, all three order-aware temporal heads outperform the position-invariant Mean baseline at every window size (Table~\ref{tab:head-results}), and the gap widens as the window grows (F1 0.918 vs.\ 0.872 at $K{=}8$; 0.896 vs.\ 0.788 at $K{=}32$). Averaged frame appearance thus carries substantial signal, but it is not enough to reliably separate intentional engagement from incidental contact such as hovering or resting a hand on an object; the temporal order of hand--object motion supplies the remaining evidence. Second, longer windows do not improve detection: TCN and BiLSTM remain roughly flat from $K{=}8$ to $K{=}32$, while the Transformer and Mean heads degrade. An 8-frame window ($\sim$267\,ms at 30\,fps) therefore appears sufficient to capture the onset of correlated motion. Combined with the low per-clip inference cost (Table~\ref{tab:efficiency}), this short window lets the detector respond shortly after a grasp-and-move begins rather than waiting for a longer gesture to complete, which is important for making opportunistic TUIs feel responsive.

\subsection{Input Modality Inference}
\label{sec:disc-summon}

The Summon results (Tables~\ref{tab:summon-features} and~\ref{tab:summon-perclass}, Figure~\ref{fig:summon-confusion}) point to two conclusions. First, input modality is conveyed mainly by how the hand moves rather than by its static pose: adding velocity and acceleration to joint positions raises generalized top-1 accuracy from 72.2\% to 80.9\%. Second, the remaining errors are structured rather than random. They concentrate in a few pairs of modalities with similar short hand motions, most notably \emph{joystick}$\leftrightarrow$\emph{switch} (short directional motions) and \emph{keypad}$\rightarrow$\emph{touchpad} (brief downward finger movements). Because the correct class in these cases is usually ranked second or third, a ranked list recovers many of them: top-3 accuracy reaches 90.8\%, and \emph{joystick} recall rises from 0.64 to 0.86. The large gain from personalization (97.5\% top-1) further suggests that much of the remaining error comes from cross-user differences in motion style rather than from modalities that are inherently indistinguishable. These findings motivate pairing top-3 suggestions with lightweight personalization. Users can recruit an object immediately by choosing from a short list, and each choice provides a labeled example; once 2--3 demonstrations per class have accumulated, the system can adapt to the individual user without a separate calibration session.

\subsection{Benefits and Costs of Opportunistic Tangibility}

The strongest interaction-level benefit of MidasTUI was physical grounding. Participants reported lower physical demand, effort, and frustration than with Mid-Air interaction, and their comments repeatedly emphasized the value of manipulating objects on a physical surface rather than sustaining unsupported movement in free space. These findings suggest that everyday objects are useful because they are readily available, provide passive haptic feedback and stable physical references during interaction.

Participants also generally perceived MidasTUI as more precise, stable, and responsive than Mid-Air interaction. This perceived sense of greater physical control suggests that tangible grounding can improve the experience of fine-grained manipulation. Participants' responses further suggest potential benefits of asymmetric bimanual interaction. Several participants preferred using different objects with the two hands and described this as reducing switching burden or helping with compound operations. Their responses suggest that opportunistic TUIs may be particularly useful when the two hands assume complementary roles.

These benefits, however, came with an interaction-time cost. Participants completed the task faster with Mid-Air than with MidasTUI. One possible explanation is that tangible interaction introduces additional acquisition and transition overhead, as users must locate, grasp, manipulate, and release physical objects when moving between tangible controls. Nevertheless, participants rated switching between tools more favorably with MidasTUI than with Mid-Air interaction, despite the need to physically acquire different objects. One possible explanation is that distinct physical objects externalize interaction state by providing spatially separated cues for different control roles, reducing the need to navigate abstract interface modes.

Overall, these results reveal a trade-off between rapid access and physically grounded control. For wearable interaction, instant tangible interfaces may be particularly valuable in tasks where comfort, stability, and sustained use outweigh the need to minimize short-term task completion time.

\subsection{TUI Interaction Design Implications}
\label{sec:disc-implications}
Our results point toward a shift in interaction design, where input is defined by user motion rather than object properties. We suggest several design implications for opportunistic TUI.

\paragraph{\textbf{Design for Physical Grounding.}}
The value of everyday objects as opportunistic TUIs is that they are always available in the environment. Our user study further suggests that their physical presence changes the interaction experience by reducing physical demand and increasing perceived stability and control. Opportunistic TUI systems should therefore favor mappings that meaningfully exploit physical support and contact, rather than treating everyday objects only as visually trackable proxies. For example, sliding or rotating an object on a tabletop may offer greater benefits than manipulating the same object in mid-air, as the surface provides a stable reference and passive haptic feedback that can support more controlled interaction. 

\paragraph{\textbf{Minimize Acquisition and Transition Overhead.}}
Although participants preferred MidasTUI on physical support and control, task completion was slower than with Mid-Air interaction. Future systems should therefore reduce the cost of recruiting and switching among physical controls. Possible strategies include persistent object-function mappings, faster reactivation of previously registered objects, reuse of a smaller number of physical objects across multiple controls, or hybrid techniques that combine tangible and Mid-Air interaction depending on the operation.

\paragraph{\textbf{Choice of Objects as TUIs.}}
Although engagement detection performed well overall, camera-based hand-object sensing remains sensitive to object properties. Small objects, for example, may yield less stable detection because they offer limited visible contact area and are more easily occluded by the hand. Other object features such as shape and graspability may also affect detection reliability. We suggest that opportunistic TUI systems do not treat all everyday objects as suitable, but instead favor objects whose physical form supports both robust visual detection and meaningful passive haptics in real practice.

\paragraph{\textbf{Choice of Supported Input Modalities.}}
Similarly, not all input modalities are equally suitable for everyday objects as TUIs, even if users can demonstrate them. Because most everyday objects are passive and rigid, they are better suited to motion-based input than to elastic or force-based interaction. In lightweight camera-based systems, force sensing is difficult, making isometric interaction less practical. By contrast, isotonic interactions based on visible translation or rotation are more compatible with both camera sensing and the physical affordances of everyday objects.

\paragraph{\textbf{Ground TUI Interaction in Hand Movement.}}
Object-to-function mappings can be grounded in the structure of the user’s hand movement rather than relying on specific object’s attributes to enable more adaptive and convenient interaction. In our system, digital functions are inferred from how the hand moves with or on the object, such as translating or rotating. This suggests that opportunistic TUI systems should prioritize motion patterns that are easy to recognize, repeat, and associate with control behavior, preserving flexibility across diverse objects.

\paragraph{\textbf{Compatibility with Ordinary Object Use.}}
Opportunistic TUIs should preserve the ordinary functions of everyday objects rather than override them. The system need to avoid treating every touched object as an input device, which would cause unintended activations and interfere with normal use. Instead, objects should transition smoothly between ordinary and TUI functions based on context and intent, for example through a designated interaction area or an explicit demonstration. This allows the same object to serve as a virtual control when deliberately recruited, while remaining available for its everyday purpose.

\subsection{Limitations and Future Work}
\label{sec:disc-limitations-future}

The generalized Summon model was evaluated with 16 participants, so its performance across a broader range of users remains to be established. Future work could expand the participant pool and investigate few-shot adaptation methods to improve cross-user generalization.

MidasTUI also relies on robust visual sensing of hand-object interaction, making it sensitive to occlusion, viewpoint changes, clutter, and object appearance. Move Together is especially weaker on small objects such as rings, keys, and SD cards, which are often heavily occluded by the hand. This suggests the current detector works better when at least part of the object remains visible during manipulation. Robustness could be improved through multimodal sensing, stronger temporal reasoning, personalized adaptation, or added cues such as hand pose or tactile signals.

Although our task set covers representative interface controls, it does not capture the full diversity of real-world productivity workflows. The current interaction vocabulary is limited to a small set of modalities, such as sliding, rotation, and discrete pressing or tapping, and does not reflect the richness of everyday object manipulation. The supported digital controls are likewise narrow compared with the range of functions in modern applications. Expanding both the detectable interaction modalities and the supported digital controls would enable opportunistic TUIs to support a broader range of productivity and creative tasks.

Our user study was designed as an exploratory evaluation of how MidasTUI is experienced in a realistic compound task. We therefore focused on participants' perceived workload, usability, control, and interaction preferences when interacting with MidasTUI and Mid-Air. The study involved 10 participants and a single photo-collage task, so the observed trade-offs may not generalize to all applications or longer-term use. Although participants generally perceived MidasTUI as more precise and stable, these results were interpreted as subjective assessments of control quality. In addition, one- and two-handed interaction were not tested as independent controlled conditions, so observations regarding bimanual interaction remain exploratory. Finally, participants used a fixed set of everyday objects in a controlled tabletop environment. Future work could extend this evaluation with more controlled studies focused on precision-demanding tasks, broader task domains, and longer-term use.

\section{Conclusion}
\label{sec:conclusion}

We presented MidasTUI, a demonstration-driven approach that turns everyday objects into tangible controls for wearable displays. Rather than relying on pre-authored mappings, the system infers input behavior from how users manipulate objects. We achieve this through two recognition models, \emph{Move Together} for hand-object engagement detection and \emph{Summon} for input modality inference, together with a TUI manager and a mapping layer that bind recognized interactions to common controls.

Our evaluations demonstrate robust performance across diverse everyday objects and previously unseen users. A comparative user study further showed that MidasTUI reduced perceived physical burden and improved usability and control relative to Mid-Air interaction, although it required longer task completion time. Together, these results demonstrate the potential of everyday objects as instant tangible interfaces for wearable displays, offering physical grounding and opportunistic interaction without additional input hardware. MidasTUI ultimately points toward a future in which wearable devices turn the surrounding physical environment into an interaction resource, with instant interface behavior emerging from users' familiar actions.


\bibliographystyle{ACM-Reference-Format}
\bibliography{src/reference}

\clearpage
\appendix

\section{Appendix}
\label{appendix}

\subsection{Research Transparency}
\label{appendix:meta}

\subsubsection{Data and Code Availability}
\label{appendix:data-release}

The training datasets for both the Move~Together classifier, which detects hand--object engagement, and the Summon classifier, which infers input modality, are available through the project's OSF repository via this \href{https://osf.io/h2gwq/overview?view_only=2dc51117cecb4a248d5abd436ea6b51a}{\textcolor{linkblue}{link}}. The repository contains: (i) the hand-object engagement detection capture used to train the classifier, and (ii) the Summon capture used to train the input-modality classifier, consisting of per-frame 3D hand-joint sequences from $16$ participants across $10$ interaction classes. 

\subsection{Model Architecture Details}
\label{appendix:arch}

\subsubsection{Move~Together Model}

The Move~Together classifier takes a clip of $K$ consecutive
RGB frames and predicts a binary engagement label.
It consists of a spatial backbone and a temporal aggregation head.

\paragraph{Spatial Backbone.}
We use FastViT-T8~\cite{vasu2023fastvit}, a hybrid vision
transformer that interleaves RepMixer token-mixing blocks
with convolutional feed-forward layers.
Each frame $I_t \in \mathbb{R}^{3 \times 224 \times 224}$ is
independently encoded to a $D$-dimensional feature vector
via global average pooling, where $D = 384$ (the native
FastViT-T8 feature dimension).
The backbone is initialised with ImageNet-pretrained weights
and fine-tuned end-to-end.
Total backbone parameters: 3.26M.

\paragraph{Temporal Heads.}
We evaluate four temporal aggregation heads
(Section~\ref{sec:eval-move-together}).
Here, $D{=}384$ denotes the backbone output dimension.
The 256-dimensional hidden representations are formed within
the heads; the BiLSTM and Transformer use input projections
from 384 to 256 dimensions.

\begin{itemize}
  \item \textbf{Mean.}
    Mean-pooling over the $K$ frame features, followed by a
    two-layer MLP ($D \to 256 \to 1$) with ReLU and dropout.
    Parameters: 0.19M (head only).

  \item \textbf{TCN.}
    Two 1-D causal convolutional layers
    (kernel size 3, $D \to 256 \to 256$) with BatchNorm,
    ReLU, and dropout, followed by a linear classifier
    on the last timestep.
    Parameters: 0.26M.

  \item \textbf{BiLSTM.}
    A linear projection with LayerNorm and GELU
    ($D \to 256$), a 2-layer bidirectional LSTM
    (hidden dim 256, output dim 512), and a per-frame
    linear classifier ($512 \to 1$) with LayerNorm
    and dropout. Unlike the other heads, the BiLSTM
    produces per-frame predictions; chunk-level output
    is obtained by averaging.
    Parameters: 2.83M.

  \item \textbf{Transformer.}
    A linear projection ($D \to 256$), learned positional
    encoding, and a 2-layer Transformer encoder
    (4~heads, feed-forward dim 1024, GELU, pre-norm)
    followed by LayerNorm and a linear classifier on the
    last position.
    Parameters: 1.80M.
\end{itemize}

\paragraph{Hand-Guided Crop.}
Input frames are cropped around the hand region using a
binary hand segmentation mask from the last frame in the clip.
The crop is a square bounding box centred on the mask centroid
with 30\% padding.
During training, random jitter (${\pm}15$\% of crop size)
and scale perturbation ($0.8$--$1.2{\times}$) are applied;
at inference, the crop is deterministic.
The same crop parameters are shared across all $K$ frames
in a clip for temporal consistency.

\subsubsection{Summon Model}

The Summon classifier takes a variable-length sequence of
26-joint hand positions and predicts one of 10 input-device
classes.

\paragraph{Input Representation.}
Each frame contains the $(x, y, z)$ world-space coordinates
of 26 tracked hand joints (78 dimensions).
Frame-to-frame velocity (78d) and acceleration (78d) are
concatenated, yielding $\mathbf{s}_t' \in \mathbb{R}^{234}$
per frame.
The first and last 10\% of each sequence are trimmed to
remove transition noise.

\paragraph{Architecture.}
We use a 1-D ResNet with the following structure:
\begin{itemize}
  \item \textbf{Stem:} Conv1D (kernel 7, padding 3) $\to$
    BatchNorm $\to$ ReLU.
    Maps 234 input channels to $C{=}64$.
  \item \textbf{Stage~0:} 2 residual blocks at $C{=}64$.
    Each block: Conv1D(3) $\to$ BN $\to$ ReLU $\to$
    Dropout $\to$ Conv1D(3) $\to$ BN + skip connection.
  \item \textbf{Stages~1--3:} Each stage doubles channels
    and halves temporal resolution via stride-2 convolution,
    followed by 2 residual blocks.
    Channel progression: $64 \to 128 \to 256 \to 512$.
  \item \textbf{Pooling:} Length-aware masked global average
    pool over valid timesteps.
  \item \textbf{Classifier:} Dropout($0.4$) $\to$ Linear($512 \to 10$).
\end{itemize}
Total parameters: 4.74M.

\subsection{Training Configuration}
\label{appendix:training}

\subsubsection{Move~Together Training}

Table~\ref{tab:mt-config} lists the training hyperparameters
for the Move~Together classifier.
All four temporal heads share the same backbone, optimiser,
and learning rate schedule; they differ only in batch size
(16 for BiLSTM due to its higher memory footprint, 32 for the
rest) and warmup duration.
We use mixed-precision training (FP16) with gradient clipping
at max norm~1.0.
Data augmentation is applied at the clip level: a hand-guided
crop with random jitter and scale perturbation ensures spatial
consistency across frames, while colour jitter and horizontal
flips are sampled once per clip and applied uniformly to all
$K$~frames for temporal coherence.
Early stopping on validation F1 with patience~10 prevents
overfitting, as AUROC degrades after epoch~10--15 despite
stable F1.

\begin{table}[h]
  \caption{Move~Together training hyperparameters.}
    \Description[Move Together training hyperparameters including optimizer, learning rate, augmentation, and preprocessing.]{A two-column table listing Move Together training hyperparameters. Key values: AdamW optimizer, learning rate 1e-4, cosine annealing schedule, batch size 32 (16 for BiLSTM), 60 epochs with early stopping patience 10, FP16 mixed precision, hand-guided crop with jitter and scale augmentation, color jitter, and ImageNet normalization.}
  \label{tab:mt-config}
  \centering
  \small
  \begin{tabular}{ll}
    \toprule
    \textbf{Hyperparameter} & \textbf{Value} \\
    \midrule
    Optimiser          & AdamW \\
    Learning rate      & $1 \times 10^{-4}$ \\
    Weight decay       & $1 \times 10^{-4}$ \\
    LR schedule        & Cosine annealing \\
    Warmup epochs      & 5 (Transformer, TCN, Mean); 3 (BiLSTM) \\
    Batch size         & 32 (Transformer, TCN, Mean); 16 (BiLSTM) \\
    Epochs             & 60 \\
    Early stopping     & Patience 10 on val F1 \\
    Mixed precision    & FP16 (PyTorch AMP) \\
    Gradient clipping  & Max norm 1.0 \\
    Input resolution   & $224 \times 224$ \\
    Window size $K$    & 8 (default; ablated over 8, 16, 32) \\
    \midrule
    \multicolumn{2}{l}{\textit{Data augmentation}} \\
    Hand-guided crop   & Jitter ${\pm}15$\%, scale $0.8$--$1.2{\times}$ \\
    Colour jitter      & Brightness/contrast/saturation ${\pm}0.2$, hue ${\pm}0.1$ \\
    Horizontal flip    & $p = 0.5$ \\
    \midrule
    \multicolumn{2}{l}{\textit{Preprocessing}} \\
    Normalisation      & ImageNet mean/std \\
    Hand mask source   & MediaPipe hand segmentation \\
    \bottomrule
  \end{tabular}
\end{table}

\subsubsection{Summon Training}

Table~\ref{tab:summon-config} lists the training
hyperparameters for the Summon classifier.
The model is trained with a higher learning rate
($10^{-3}$ vs.\ $10^{-4}$ for Move~Together) since
the ResNet1D head has no pretrained backbone.
Label smoothing (0.1) and Mixup ($\alpha{=}0.2$) are
applied to regularise against the small per-class sample
size (${\sim}$84 training sequences per class in LOPO).
Temporal augmentation includes random cropping (up to 30\%
of frames removed), Gaussian noise injection on joint
coordinates, scale jitter, and time reversal.
The first and last 10\% of each sequence are trimmed before
feature extraction to remove transition artefacts at the
start and end of demonstrations.
Early stopping on validation accuracy with patience~100
accommodates the slower convergence of the 1-D ResNet
on variable-length sequences.

\begin{table}[h]
  \caption{Summon training hyperparameters.}
    \Description[Summon training hyperparameters including optimizer, learning rate, regularization, and augmentation.]{A two-column table listing Summon training hyperparameters. Key values: AdamW optimizer, learning rate 1e-3, cosine annealing, batch size 16, up to 1000 epochs with early stopping patience 100, label smoothing 0.1, Mixup alpha 0.2, temporal crop up to 30 percent, Gaussian noise 0.005, scale jitter 5 percent, time reversal, and 10 percent trim ratio.}
  \label{tab:summon-config}
  \centering
  \small
  \begin{tabular}{ll}
    \toprule
    \textbf{Hyperparameter} & \textbf{Value} \\
    \midrule
    Optimiser          & AdamW \\
    Learning rate      & $1 \times 10^{-3}$ \\
    Weight decay       & $1 \times 10^{-4}$ \\
    LR schedule        & Cosine annealing \\
    Warmup epochs      & 5 \\
    Batch size         & 16 \\
    Max epochs         & 1000 \\
    Early stopping     & Patience 100 on val accuracy \\
    Mixed precision    & FP16 (PyTorch AMP) \\
    Label smoothing    & 0.1 \\
    Mixup alpha        & 0.2 \\
    Loss               & Cross-entropy \\
    Evaluation         & LOPO (16 participants) \\
    \midrule
    \multicolumn{2}{l}{\textit{Data augmentation}} \\
    Temporal crop      & Remove up to 30\% of frames (random) \\
    Gaussian noise     & $\sigma = 0.005$ \\
    Scale jitter       & ${\pm}5$\% \\
    Time reversal      & $p = 0.5$ \\
    \midrule
    \multicolumn{2}{l}{\textit{Preprocessing}} \\
    Trim ratio         & 10\% removed from start and end \\
    Normalisation      & Per-dataset mean/std (training split) \\
    Features           & pos + velocity + acceleration (234d) \\
    \bottomrule
  \end{tabular}
\end{table}

\subsection{Hand Motion Trajectory Visualisation}
\label{appendix:trajectories}

\begin{figure*}[t]
  \centering
  \Description[Grid of overlaid 3D hand joint trajectories for 16 participants across 10 interaction classes.]{Rows are labeled p1 through p16. Columns, from left to right, are slider, knob, switch, touchpad, joystick, mouse, pen, d-pad, keypad, and pointer. Each cell overlays the trajectories of all 26 tracked hand joints across the demonstration sequences for that participant and class; color encodes temporal progression. Slider demonstrations tend to form elongated sweeps, knob demonstrations form curved traces, and keypad demonstrations form compact tapping patterns. The orientation, extent, and shape vary across participants, with visibly different switch and joystick trajectories illustrating cross-user variation.}
  \includegraphics[width=0.98\textwidth,height=0.82\textheight,keepaspectratio]{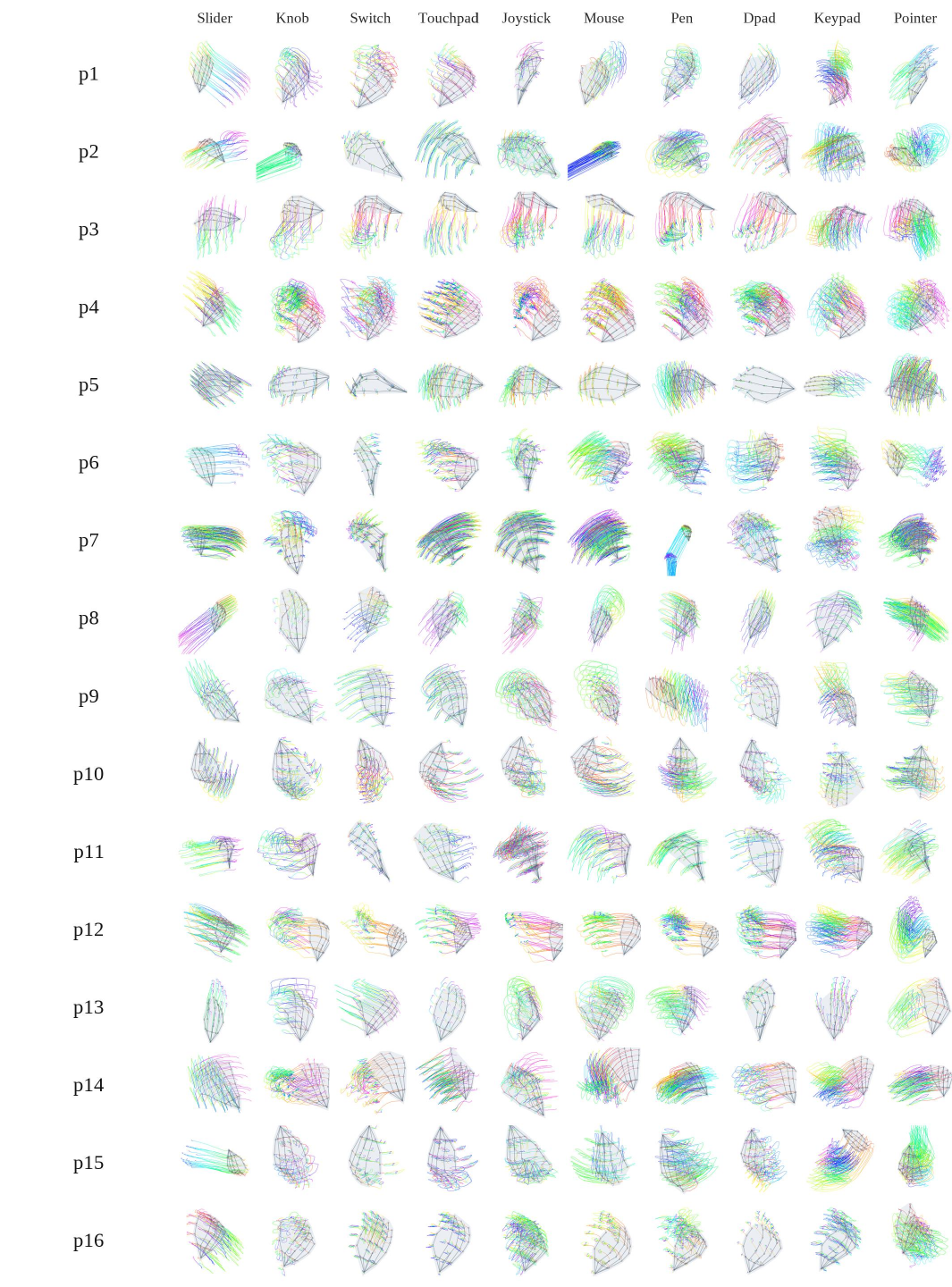}
  \caption{Overlaid 3D hand joint trajectories for 16~participants
    $\times$ 10~interaction classes. Each cell visualises all
    demonstration sequences for one participant--class pair.
    Classes exhibit distinct motion signatures (e.g., the linear
    sweep of \emph{slider}, the circular trace of \emph{knob}),
    but the same class can look markedly different across
    participants---particularly for \emph{switch} and
    \emph{joystick}---explaining the cross-user confusion
    patterns in Figure~\ref{fig:summon-confusion}.}

  \label{fig:summon-montage}
\end{figure*}

Figure~\ref{fig:summon-montage} visualises the 3D hand joint
trajectories recorded during Summon demonstrations for all
16~participants across all 10~interaction classes.
Each cell overlays the trajectories of all 26 tracked hand joints
across all demonstration sequences for that participant--class pair,
with colour encoding the temporal progression.
The visualisation reveals both the characteristic motion signatures
of each class (e.g., the linear sweep of \emph{slider}, the
circular trace of \emph{knob}, the compact tap of \emph{keypad})
and the substantial cross-participant variability that makes
leave-one-participant-out generalisation challenging---the same
interaction class can produce visually distinct trajectory patterns
across different users.

\end{document}